\documentclass[%
 reprint,
 amsmath,amssymb,
 aps,
]{revtex4-2}

\usepackage{graphicx}% Include figure files
\usepackage{dcolumn}% Align table columns on decimal point
\usepackage{bm}% bold math
\usepackage{placeins}

\begin{document}

\preprint{APS/123-QED}

%\title{Preparation-dependent shear-jamming packing fractions of elliptical particles under oscillatory shear flows}% Force line breaks with \\
\title{Orientational arrest in dense suspensions of elliptical particles under oscillatory shear flows}%


\author{Zakiyeh Yousefian}
\author{Martin Trulsson}
\affiliation{
Theoretical Chemistry, Lund University
}%

             
\begin{abstract}
We study the rheological response of dense suspensions of elliptical particles, with an aspect ratio equal to 3, under oscillatory shear flows and imposed pressure by numerical simulations. Like for the isotropic particles, we find that the oscillatory shear flows respect the Cox-Merz rule at large oscillatory strains but differ at low strains, with a lower viscosity than the steady shear and higher shear jamming packing fractions. However, unlike the isotropic cases (\emph{i.e.,}~discs and spheres), frictionless ellipses get dynamically arrested in their initial orientational configuration at small oscillatory strains. We illustrate this by starting at two different configurations with different nematic order parameters and the average orientation of the particles. Surprisingly, the overall orientation in the frictionless case is uncoupled to the rheological response close to jamming, and the rheology is only controlled by the average number of contacts and the oscillatory strain. Having larger oscillatory strains or adding friction does, however, help the system escape these orientational arrested states, which are evolving to a disordered state independent of the initial configuration at low strains and ordered ones at large strains.
\end{abstract}

\maketitle

\noindent \emph{Introduction -}
Mechanical properties of soft materials and complex fluids, such as suspensions in concentrated regimes, emulsions and granular materials, are challenging to describe due to their often complicated and dynamical many-body effects. Such systems have appeared as an important field of study not only from a pure physics perspective but also due to their practical applications in materials science and numerous other areas \cite{liu2020acoustic,kleman2007soft}. Oscillatory shear flows have been broadly used to investigate dense suspensions mechanical properties \cite{ferry1980viscoelastic,dealy2012melt,marenne2017nonlinear}. Previous studies have shown that the rheology of dense suspensions under oscillatory shear do not necessarily follow Cox-Merz rule, which states that the oscillatory shear and steady shear viscosities should be equal to each other \cite{guazzelli2018rheology}. At constant shear-rate $\dot\gamma$ the viscosity $\eta$ of dense non-Brownian suspensions consisting of rigid particles diverges as $\eta/\eta_f \sim (\phi_c-\phi)^{-\alpha}$, where $\eta_f$ is the viscosity of the background fluid, $\phi$ is the particles packing fraction and $\alpha$ is a positive exponent typically close to $2$ \cite{andreotti2012shear,guazzelli2018rheology}. The shear jamming packing fraction $\phi_c$ is dependent on various parameters, including particles shape \cite{salerno2018effect,marschall2019orientational,nagy2017rheology,donev2004improving, marschall2019shear,trulsson2018rheology,azema2015internal,brown2011shear}, friction \cite{trulsson2017effect,silbert2010jamming,seto2013discontinuous} and interactions \cite{dong2020unifying,irani2014impact,berger2016scaling,singh2019yielding}. The rheology of dense suspensions can be quite complicated at unsteady shear conditions, showing non-trivial transient rheological behaviour \cite{blanc2011local,peters2016rheology,ness2016two}. It has been shown that oscillatory shear perpendicular \cite{lin2016tunable, ness2017oscillatory,ness2018shaken} or parallel to the primary constant shear \cite{dong2020transition} can help flowability of dense suspensions by reducing the viscosity in a controlled way \cite{ness2018shaken,dong2020transition}. The viscosity reduction in suspensions is generally attributed to a restructuring of the microstructure \cite{dong2020transition} which, for instance, at orthogonal shear flows, happens through a tilting and ultimate breakage of the force chains \cite{lin2016tunable} or more generally through random organization \cite{pine2005chaos,ness2018shaken,corte2008random}. These studies were done on isotropic granules, although in reality, particles have some anisotropy.\\

\noindent The present study will explore the effect of shape anisotropy on the rheology of dense non-Brownian suspensions under oscillatory shear. \\

\begin{figure*}
\centering
\includegraphics[scale=0.7]{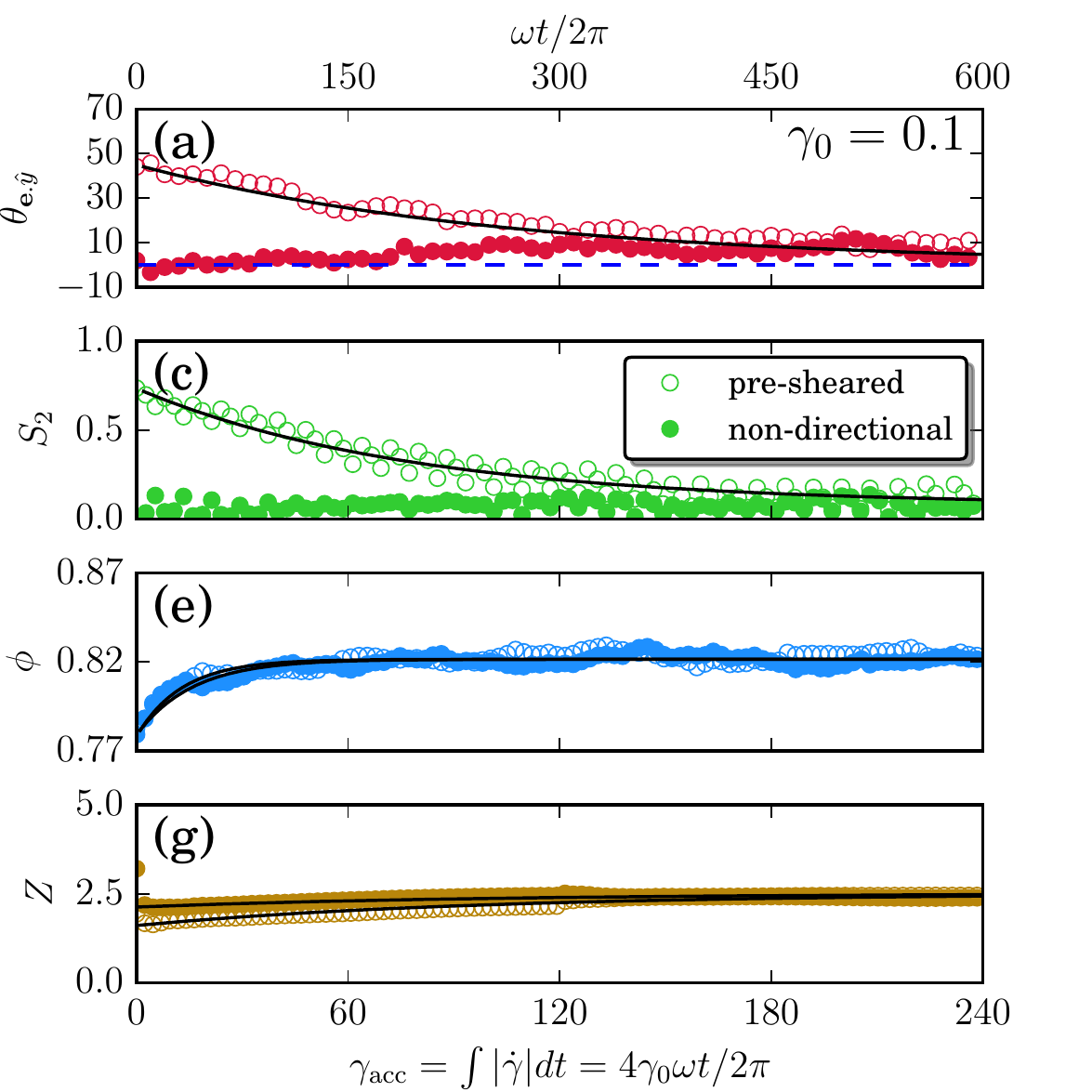}
\includegraphics[scale=0.7]{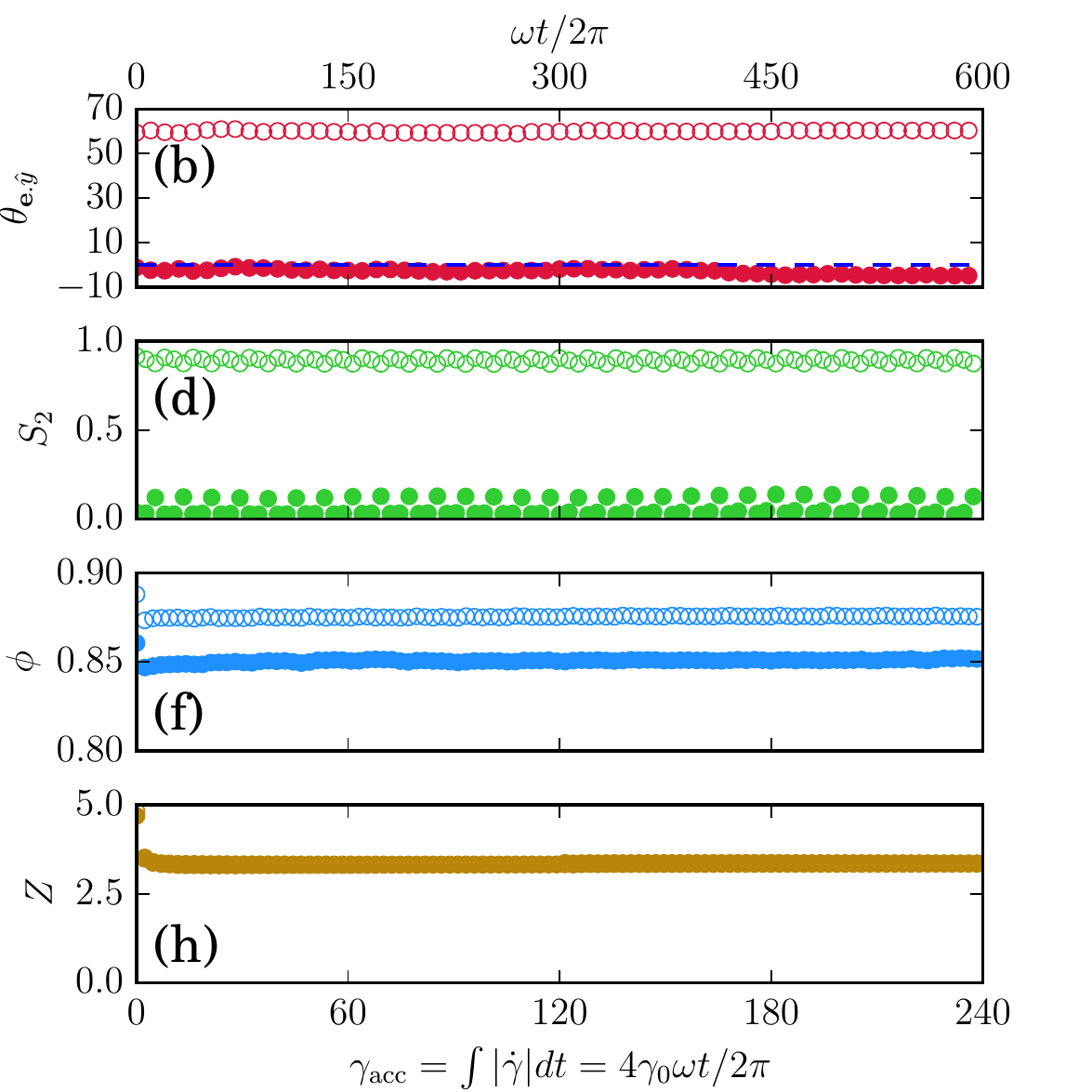}

\caption{Evolution of {\bf (a,b)} the direction angle $\theta_{\bold e \cdot \hat{y}}$($^{\circ}$), {\bf (c,d)} the nematic ordering $S_2$, {\bf (e,f)} the packing fraction $\phi$ and {\bf (g,h)} the number of contacts $Z$ for the frictional $(\mu_p=0.4)$ (left column) and frictionless $(\mu_p=0)$ (right column) configurations at $\gamma_0=0.1$ and $J'\simeq 0.1$. Empty and full symbols correspond to pre-sheared and non-directional preparations, respectively. Black lines are best fits of the relaxations. The blue dashed lines in {\bf (a,b)} indicate zero lines.}  
\label{fgr:relax_time}
\end{figure*}

\noindent \emph{Simulation Method -} Using a discrete element method we study two dimensional suspensions consisting of $\sim 1000$ ellipses with $\pm 50 \%$ polysdispersity in major axis with a flat distribution and an aspect ratio of $\alpha=a/b=3$, where $a$ and $b$ are the major and minor radius, respectively. The eccentricity of an ellipse is given as $e=\sqrt{(1-\alpha^{-2})}$. We apply a constant external pressure $\mathit{P}^{\rm ext}$ on the two rough confining walls in their normal direction (now denoted $y$-direction). The two walls have a relative oscillatory velocity difference in transverse $x$-direction which leads to a macroscopic shear rate $\dot{\gamma}(t)=\dot{\gamma}_{0}\cos(\frac{\omega}{2\pi} t)$, where $\dot{\gamma}_{0}$ is the amplitude and $\omega/{2\pi}$ is the frequency of the oscillatory shear. Wall particles have the same properties as the flowing ellipses. The corresponding strain will be ${\gamma}(t)={\gamma}_{0}\sin(\frac{\omega}{2\pi} t)$, where ${\gamma}_{0}=2\pi\dot{\gamma}_{0}/\omega$ is the amplitude of the oscillatory strain. Particles interact with each other via harmonic forces, $\mathbf{f}_{ij^{}}=\mathit{k_{n}^{}{\delta}_{n}^{ij}}\mathbf{n}^{\mathit{ij}}+\mathit{k_{t}^{}{\delta}_{t}^{ij}}\mathbf{t}^{\mathit{ij}}$, where $\mathit{k_{n}}$ and $\mathit{k_{t}}=\mathit{k_{n}}/2$ are the normal and the tangential spring constants, respectively, and where $\mathit{{\delta}_{n}^{ij}}$ and $\mathit{{\delta}_{t}^{ij}}$
are the normal overlap and tangential displacement between the particles $i$ and $j$ \cite{trulsson2018rheology}. The particles also experience torques $\tau_{ij}$. To satisfy the rigid body assumption, the ratio between the normal spring constant and the external pressure $\mathit{k_{n}}/\mathit{P}^{\rm ext}$ is set to be $3\times10^4$. Coulomb friction $|\mathbf{f}_{t}^{ij}|\leq \mu_{\mathit{p}}|\mathbf{f}_{n}^{ij}|$, is the constraint for the tangential force with $\mu_p$ being the friction coefficient of the particles which can be either $\mu_p=0$ (frictionless) or $\mu_p=0.4$ (frictional) unless otherwise stated. The force, $\bold{f}$, and torque, $\tau$, imposed by the background fluid on each elliptical particle $i$ are described by $\mathbf{f}_{i}^{\rm visc}= 3\pi\eta_{f}[c_{fa}(\mathbf{u}_{f}^{a}(y)-\mathbf{u}_{i}^{a})+c_{fb}(\mathbf{u}_{f}^{b}(y)-\mathbf{u}_{i}^{b})]$ and $\mathbf{\tau}_{i}^{\rm visc}=4\pi\eta_{f}ab[(c_{Ma}|e_{i,x}|^2+c_{Mb}|e_{i,y}|^2)\mathbf{\omega}^{f} - c_{Mr}\mathbf{\omega}_i]$ \cite{chwang1975hydromechanics,datta1999stokes}, where $\mathbf{u}_i$ and $\omega_i$ are the translational and angular velocity of the particle $i$, $\mathbf{e}_{i}=(e_{i,x},e_{i,y})$ is its unit direction vector along the major axis and  $\mathbf{u}^{a}_i=(\mathbf{u}_i.\mathbf{e}_{i})\mathbf{e}_{i}$ and $\mathbf{u}^{b}_i=\mathbf{u}_i-\mathbf{u}^{a}_i$ are the particle's velocity in major and minor axes directions, respectively.
$\mathbf{u}^f(y)=({\dot\gamma}y,0)$ is the fluid velocity with $y$ as the $y$-coordinate of the ellipse $i$ and $\mathbf{\omega}^f=\dot\gamma/2$ is the fluid angular velocity. Coefficients are $c_{fa}=\frac{8}{3}e^3[-2e+(1+e^2) \log(\frac{1+e}{1-e})]^{-1}$, $c_{fb}=\frac{16}{3}e^3[2e+(3e^2-1) \log(\frac{1+e}{1-e})]^{-1}$, $c_{Ma}=c_{fa}$, $c_{Mb}=(1-e^2)^{-1}c_{fa}$, and $c_{Mr}=\frac{4}{3}e^3(\frac{2-e^2}{1-e^2})[-2e+(1+e^2)\log(\frac{1+e}{1-e})]^{-1}$. 

\begin{figure}
\centering
\includegraphics[scale=0.7]{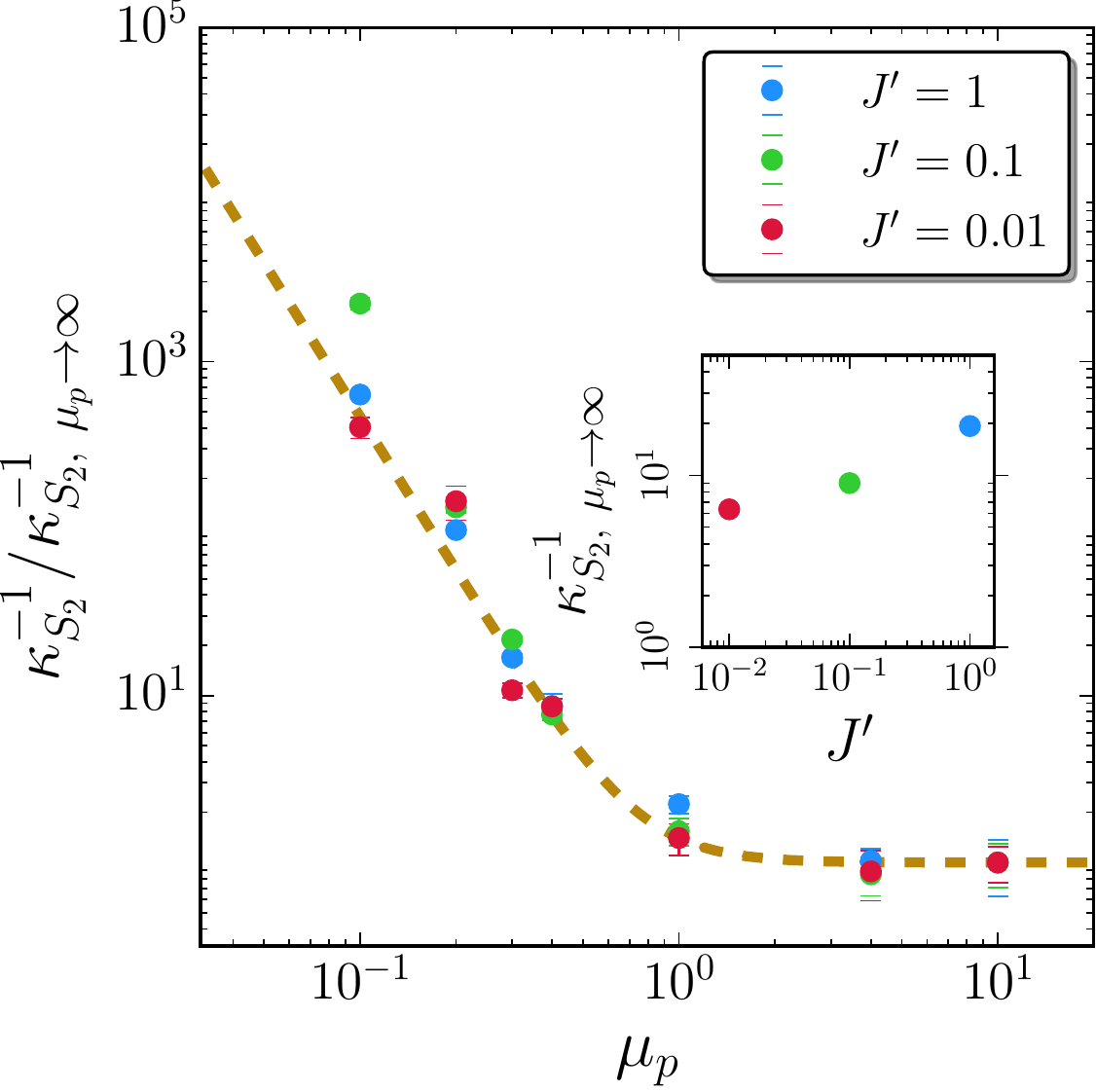}
\caption{Rescaled relaxation parameter of the nematic order $\kappa_{S_2}^{-1}$ for the pre-sheared configuration as function of $\mu_p$ at $\gamma_0=0.1$ for various $J'$. Each $J'$-curve has been normalised by its corresponding value at $\mu_p \to \infty$.. Dashed line indicates best fit of the data with $\kappa_{S_2}^{-1}\sim a_{S_2}\mu^{-\beta}_p+c$.} 
\label{fgr:relax_mu}
\end{figure}

\noindent The dynamics of the ellipses are overdamped which lead to force and torque equations given as $\mathbf{f}_{i}^{\rm ext}+\mathbf{f}_{i}^{\rm visc}= -\sum_{j}^{}\mathbf{f}_{ij}$ and $\mathbf{\tau}_{i}^{\rm ext}+\mathbf{\tau}_{i}^{\rm visc}= -\sum_{j}^{}\mathbf{\tau}_{ij}$, where $\mathbf{f}_{i}^{\rm ext}$ and $\mathbf{\tau}_{i}^{\rm ext}$ are the external forces and torques, respectively. The characteristic time is given by $t_0=3 \pi \eta_f c_{fa}/(k_n\sqrt{\alpha})$. The equation of motions are integrated by using $\delta t/t_0=0.1$ and the Heuns method. \\
We select two preparation protocols. The first one uses a pre-sheared protocol, where suspensions are first steadily sheared, and once they reached a steady state, we turn off the shear rate and let the suspension settle under the external pressure. This lead to a well-defined orientation of the ellipses. The second one corresponds to a non-directional/random protocol. We randomly placed and oriented ellipses in a dilute regime and then subjected them to compression by our external pressure. \\
Average properties are measured after the suspensions come to a steady cyclic state; that is, the averages of each oscillation period only fluctuate around particular mean values. The lowest accumulative strain for which averaged data is collected is $10$ ($\gamma_\mathrm{acc}=\int|\dot{\gamma}| dt\geq 10$) and for the largest $\gamma_0$ a minimum of one full oscillation period is considered. Within each oscillation, $60$ measurements are made (having $(2 \pi/\omega)>60 \delta t$). We excluded the $5$ closest layers to each wall in our measures to remove possible boundary effects. The stresses tensor is calculated according to $\sigma^{kl} = 1/(2A)\sum_{i<j} \mathbf{f}_{ij}^k \mathbf{r}_{ij}^l$, where $A$ is the area over which the stress is measured and $r_{ij}$ is two particles center-to-center vector. The nematic order parameter $S_2$ is taken as the largest eigenvalue of the director tensor $Q^{kl}=1/N \sum_{i} (2\mathbf{e}_i^k \mathbf{e}_i^l -\delta_{kl})$, where $N$ is the number of particles over which the measure is taken and $\delta_{kl}$ the Kronecker delta. $\theta_{\bold e \cdot \hat{y}}$ is measured as the average particle angle with respect direction $\bf \hat{y}$, normal to lower the surface. 
\\

\noindent\emph{Suspensions oscillatory-shear rheology -} 
Despite the steady shear rheology with solely viscous stresses and viscosities, at pure oscillatory shear flows, the stress and the viscosity will be a combination of both elastic and viscous responses. The complex stress in the linear response regime is given as \cite{marenne2017nonlinear,ishima2020scaling} $\sigma(t)=\eta^{\prime\prime}\dot\gamma_0\sin(\omega t)+\eta^{\prime}\dot\gamma_0\cos(\omega t)$, where $\eta^{\prime\prime}$and $\eta^{\prime}$ are the corresponding elastic (imaginary) and viscous (real) part of the complex viscosity $\eta^\ast=\eta^{\prime}-i\eta^{\prime\prime}$ 
and $t$ the time. The magnitude of the complex viscosity is given by $|\eta^\ast|=\sqrt{\eta^{\prime2}+\eta^{\prime\prime2}}$, with $\eta^{\prime}=\frac{\int_{0}^{2\pi/\omega} \sigma(t)\cos(\omega t) \, dt}{\dot\gamma_0\int_{0}^{2\pi/\omega} \cos^2(\omega t) \, dt}$ and $\eta^{\prime\prime}=\frac{\int_{0}^{2\pi/\omega} \sigma(t)\sin(\omega t) \, dt}{\dot\gamma_0\int_{0}^{2\pi/\omega} \sin^2(\omega t) \, dt}$ \cite{dong2020transition}. In a similar manner a shear-rate averaged viscous number $J'$ is defined as $J'=\frac{\eta_{f}\int_{0}^{2\pi/\omega} (\dot\gamma/P)\dot{\gamma}(t) \, dt}{\int_{0}^{2\pi/\omega} |\dot{\gamma}| \, dt}$ \cite{dong2020transition} (See SI for definition of shear-rate-averaged $\phi$, $Z$, $S_2$, and viscous/elastic stress ratio $\mu=\sigma/P$).
\\
\begin{figure}
\centering
\includegraphics[scale=0.385]{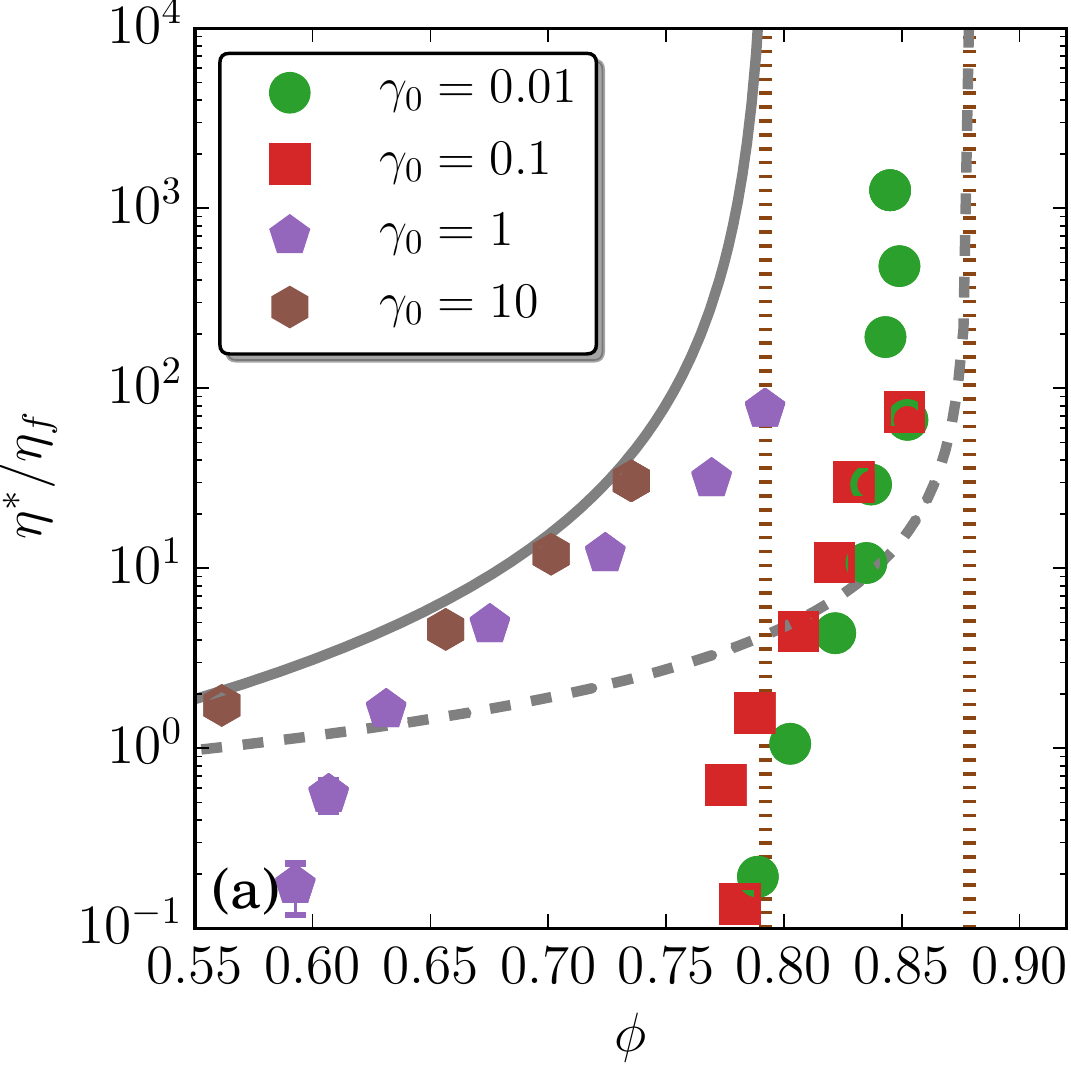}
\includegraphics[scale=0.385]{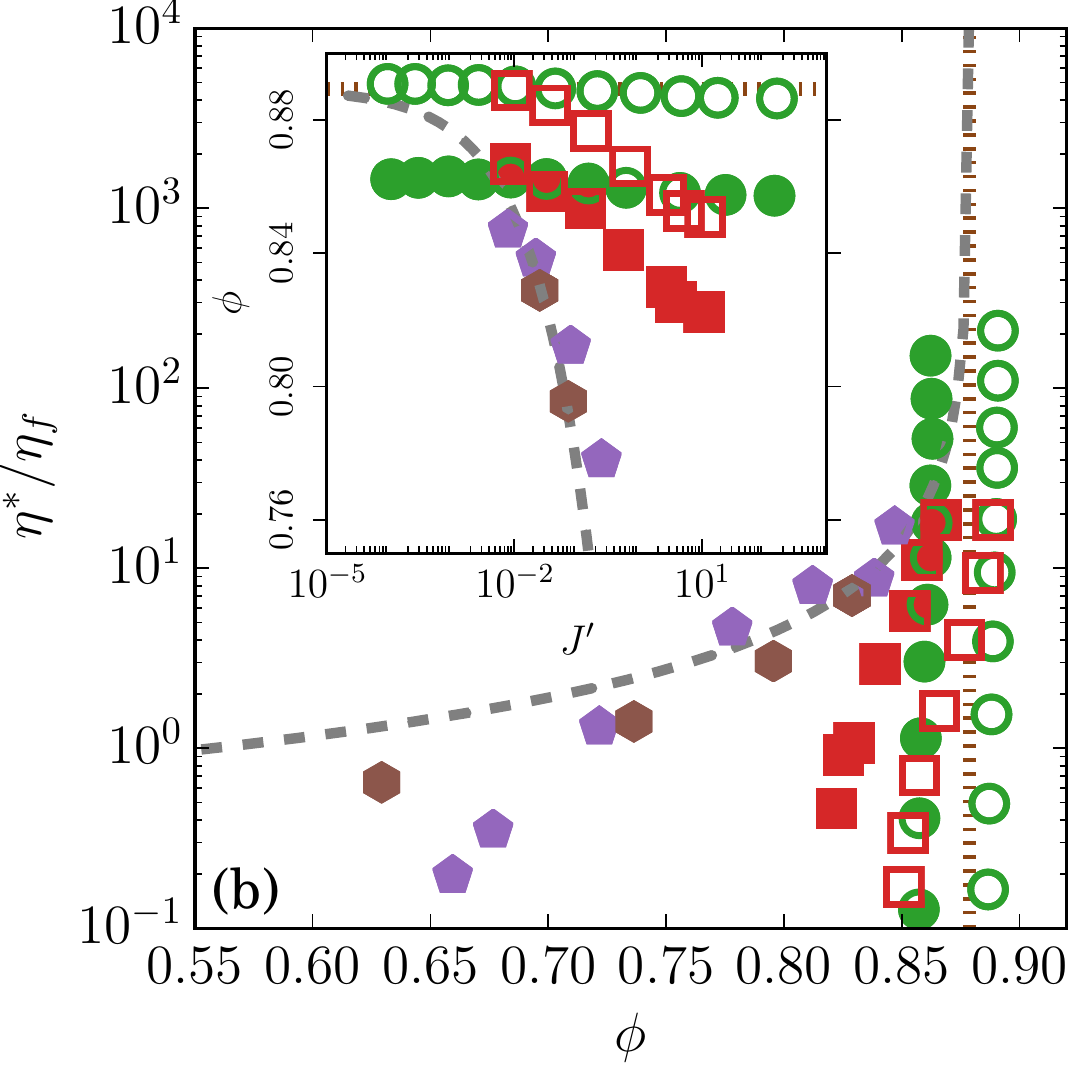}

\includegraphics[scale=0.385]{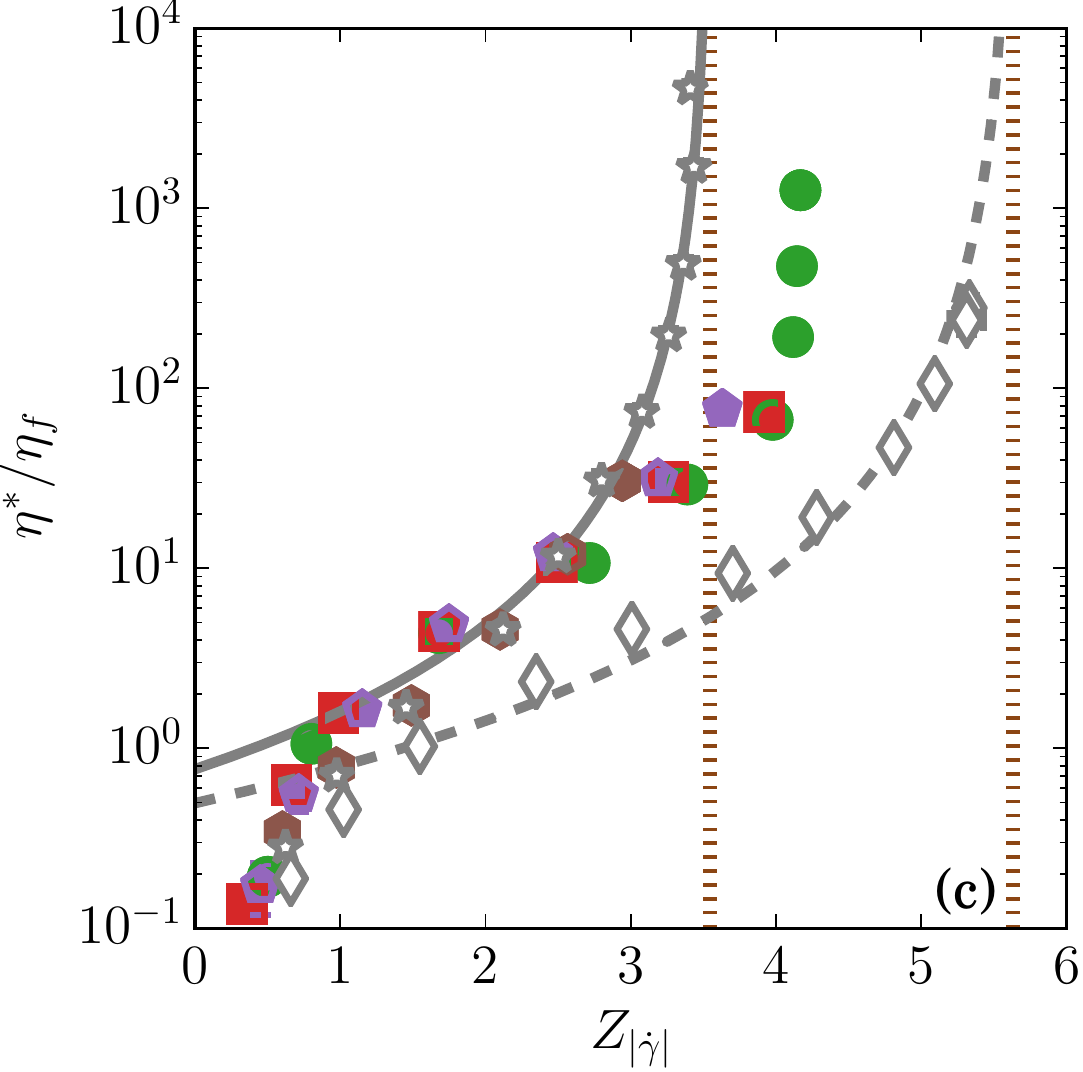}
\includegraphics[scale=0.385]{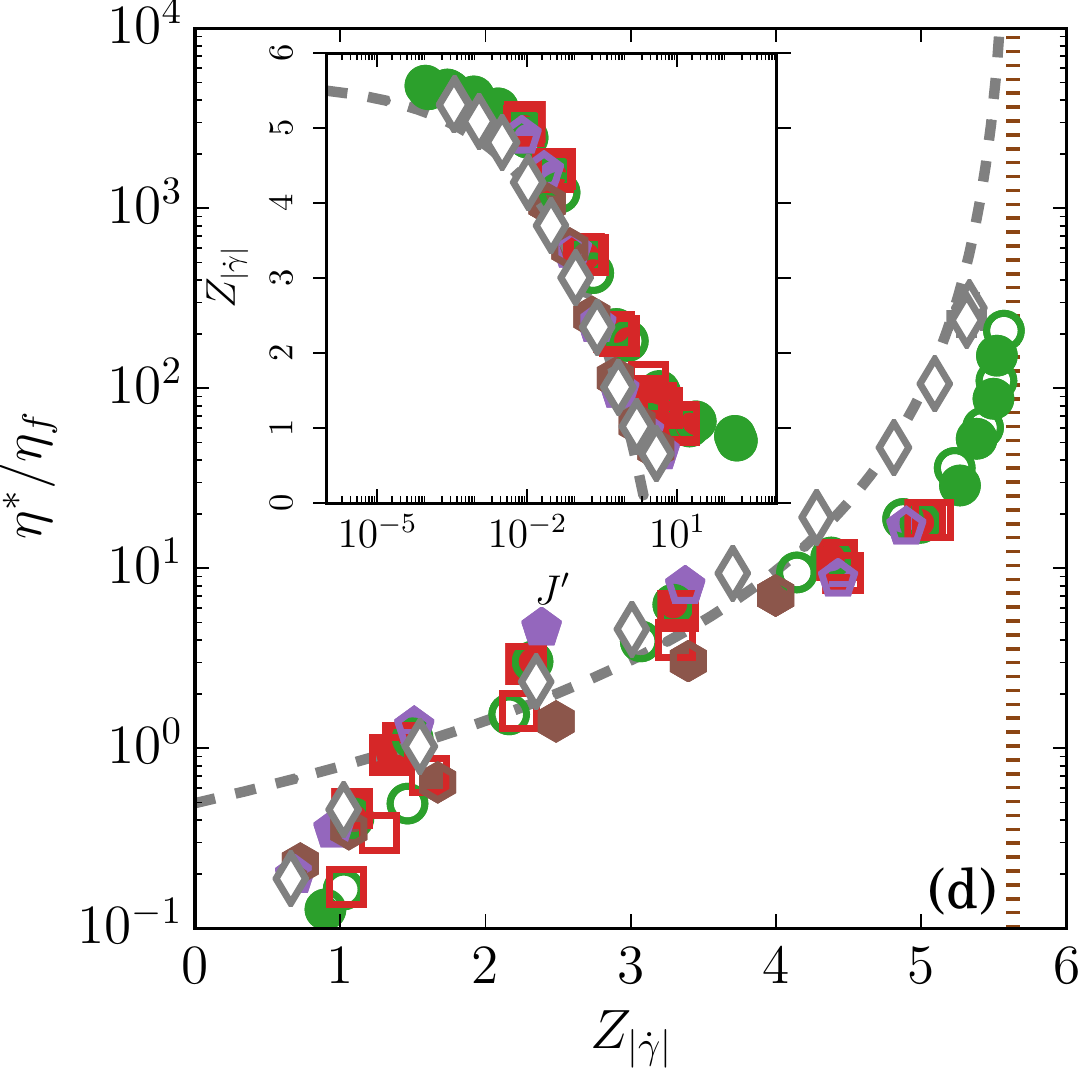}

\caption{Normalized complex viscosities $|\eta^\ast|/\eta_f$ versus the packing fraction $\phi$ ({\bf (a)} and {\bf (b)}) and the number of contacts $Z_{|\dot\gamma|}$ ({\bf (c)} and {\bf (d)}) at various strain magnitudes $\gamma_0$. {\bf (a)} and {\bf (c)} belong to frictional particles while {\bf (b)} and {\bf (d)} express the frictionless. Empty and full symbols correspond to pre-sheared and non-directional preparations, respectively. The grey solid and dashed lines are the corresponding steady shear viscosity curves for frictional and frictionless particles, respectively. The brown dotted vertical lines in {\bf (a)} and {\bf (b)} show the steady shear jamming packing fractions for frictional $\phi_{c,\mathrm{f}}^\mathrm{SS}$ and frictionless $\phi_{c,\mathrm{nf}}^\mathrm{SS}$ suspensions while in {\bf (c)} and {\bf (d)}, they indicate the steady shear jamming number of contacts for frictional $Z_{c,\mathrm{f}}^\mathrm{SS}$ and frictionless $Z_{c,\mathrm{nf}}^\mathrm{SS}$ configurations.
The inset in figure {\bf (b)} shows $\phi$ as a function of $J'$ for the frictionless ellipses, where the grey dashed line indicates the corresponding steady shear curve. Similarly, the inset in figure {\bf (d)} illustrates $Z$ versus $J'$ for the non-frictional particles, with the dashed line showing the respective steady shear case.}   
\label{fgr:viscosity}
\end{figure}

\begin{figure*}
\centering
\includegraphics[scale=0.8]{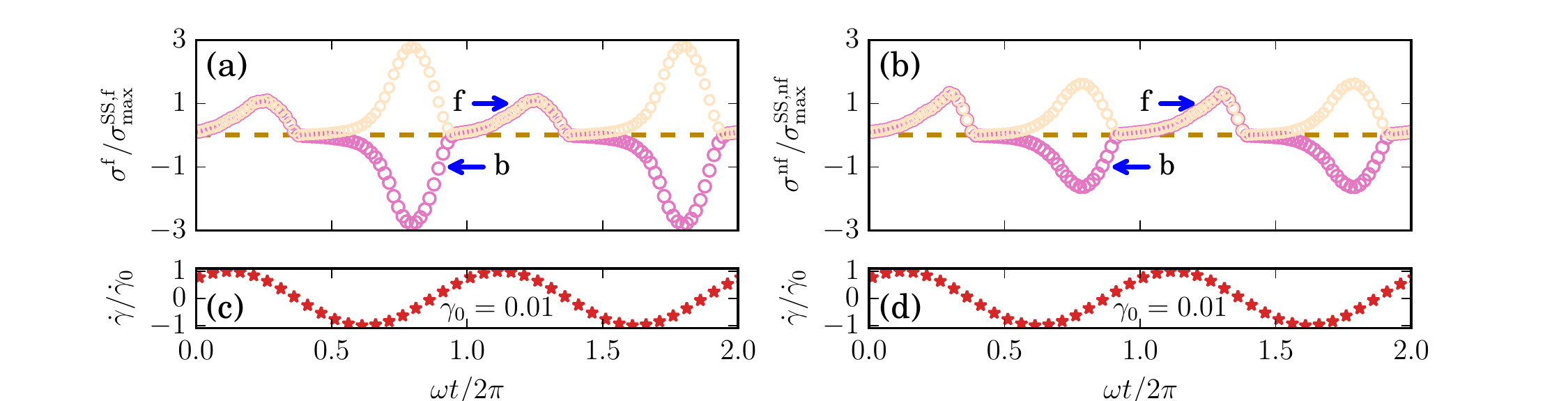}
\caption{Time series of the normalized shear stress (pink symbols) for {\bf (a)} frictional $\sigma^{\mathrm{f}}/\sigma^\mathrm{SS,f}_\mathrm{max}$, and {\bf (b)} frictionless $\sigma^{\mathrm{nf}}/\sigma^\mathrm{SS,nf}_\mathrm{max}$ pre-sheared configurations at $\gamma_0=0.01$ and $J'\simeq0.1$. $\sigma^\mathrm{SS,f}_\mathrm{max}$ and $\sigma^\mathrm{SS,nf}_\mathrm{max}$ are the stresses at the corresponding $J'$ in steady shear for frictional and frictionless ellipses, respectively. {\bf f} and {\bf b} in the figures stand for forward and backward shearing, \emph{i.e.,} along with the average orientation of the particles and opposite to it, correspondingly. The pale bisque profiles show absolute magnitudes of the shear stress. Figs.~{\bf (c)} and {\bf (d)} show the corresponding rescaled shear-rates $\dot\gamma/\dot\gamma_{0}$.}
\label{fgr:stress_forwardback}
\end{figure*}

\noindent\emph{Results and Discussion -} Fig.~\ref{fgr:relax_time} shows some typical numerical time-evolutions of $\theta_{\bold e \cdot \hat{y}}$, $S_2$, $\phi$ and Z for oscillatory shear (OS) from the two preparation protocols, pre-sheared and random packings, at a low oscillatory strain ($\gamma_0=0.1$) for both frictional and frictionless particles. The packings are initially at rest. 
For the frictional case, the two starting configurations converge to the same values, characterised by a packing fraction higher than the corresponding steady-shear value at the same viscous number and a directional disordered state with a low nematic ordering and average orientation fluctuating around zero. These orientational values are close, if not identical, to the initial random configuration, \emph{i.e.,}~the pre-sheared samples relax to random configurations. One also notices that the packing fraction relaxes much faster than the two direction parameters. The number of contacts also relaxes quite rapidly and become identical for the two protocols. On the other hand, interestingly, this is not the case for the frictionless ellipses, where all quantities except the number of contacts of the two preparation protocols stay separated and without any detectable relaxation within our numerical time window.\\ 
To get further insight, we estimate strains over which the the various quantities possibly relax. We assume exponential relaxations as $A(\gamma_\mathrm{acc})=A_\infty+(A_0-A_\infty)\exp(-\kappa \gamma_\mathrm{acc})$, where $A$ can be either of $\theta_{\bold e \cdot \hat{y}}$, $S_2$, $\phi$ or $Z$, and $\kappa^{-1}$ a relaxation strain. Fig.~\ref{fgr:relax_time} shows that the relaxation matches well an exponential, shown by solid black lines, in the frictional case. Generally, the relaxations of $S_2$ and $\theta_{\bold e \cdot \hat{y}}$ follow each other (see Fig.~\ref{fgr:relax_time}$(a)$ and $(c)$ and SI), while $\phi$ and $Z$ relax an order of magnitude faster than the two other quantities (see Fig.~\ref{fgr:relax_time}$(e)$). The relaxation strains depend on friction coefficient, oscillatory strain (see SI), and $J'$. According to Fig.~\ref{fgr:relax_time} at a typical small $\gamma_0$ (\emph{e.g.,}~0.1), the packing fraction relaxes almost instantaneously for purely frictionless particles, but to different values depending on if one starts from a random or pre-sheared configuration. The orientational relaxation from a pre-sheared configuration does, however, show a seemingly infinite relaxation strain. Fig.~\ref{fgr:relax_mu} shows that the relaxation strains in $S_2$ possibly diverge as $\kappa_{S_2}^{-1} \sim \mu_p^{-\beta}$ with $\beta=3.1\pm0.3$. Due to its sharp divergence, it is hard to tell if such divergence occurs at a finite $\mu_p$ rather than $\mu_p=0$ (see SI). Nevertheless, frictionless ellipses do not relax for small oscillatory strains. Adding a finite amount of friction does, however, help the system to relax the system's orientation, a process similar to ratcheting \cite{Alonso2004rachet}. Above $\mu_p=1$, the relaxation strains saturate and only depend on $J'$ and $\gamma_0$. Smaller $J'$ values display smaller relaxations (in terms of strains), reflecting the higher packing fractions in those cases, with a larger number of contacts and number of collisions per strain. \\
Similarly we find that these infinite relaxation strains for the frictionless appears first when $\gamma_0<0.3$, with a power-law divergence $\kappa_{S_2}^{-1} \sim (\gamma_0-\gamma_{0,c})^{-\nu}$, with $\nu\sim 3.6$ and $\gamma_{0,c} \simeq 0.1$ (see SI). \\
Further evidence that frictional ellipses relaxe but  frictionless not can be found when studying the transient stress response. Fig.~\ref{fgr:stress_forwardback} shows the stress response of pre-sheared samples after an initial relaxation in packing fraction. The frictional case shows a clear asymmetric response when sheared along the pre-sheared direction compared to opposite to it, while the frictionless case lacks this (within the noise, see SI for more details). This asymmetry leads to an imbalance in dissipation and elastic storage in the various direction that helps to evolve the system. While this asymmetric stress response is only transient, it persists up to 100 accumulated strains (and much larger than the typical turnover strain), corresponding to \emph{e.g.,~} 2700 oscillations at $\gamma_0=0.01$ at $J'\simeq0.1$. \\
Having determined that the frictionless particles do not fulfil one unique equation of state for low oscillatory strains, we investigate how the complex viscosity varies with packing fraction, and $\gamma_0$ for our two preparation protocols compared to frictional particles, see Fig.~\ref{fgr:viscosity} $(a,b)$. For both $(a)$ frictional and $(b)$ frictionless ellipses, we find that the rheological response (\emph{i.e.,}~$|\eta^*|$ vs.\ $\phi$) behave as their corresponding steady shear (SS) cases for large $\gamma_0$'s. At lower $\gamma_0$'s, one finds a lower viscosity than SS at the same packing fraction, with an increased shear jamming packing fraction for the frictional particles \emph{i.e.,} $\phi_{c,\mathrm{f}}^\mathrm{SS}<\phi_{c,\mathrm{f}}^\mathrm{OS}$. These findings are in line with what have previously been reported for discs \cite{dong2020transition,dong2020oscillatory} and spheres \cite{lin2016tunable,ness2018shaken}.
Unlike for isotropic particles, fricionless ellipses show a preparation dependent rheology (compare full and open symbols), with higher shear jamming packing fractions for pre-sheared preparations compared to non-directional/random ones. For the fricitonless ellipses the non-directional preparation yields a shear-jamming below its corresponding point in SS (\emph{i.e.,}~$\phi_{c,\mathrm{nf}}^\mathrm{OS,ran}<\phi_{c,\mathrm{nf}}^\mathrm{SS}$).
This shows that there exist at least two well-separated oscillatory shear-jamming points for frictionless elliptical particles. Nonetheless, once the complex viscosity $|\eta^*|/\eta_f$ is plotted against the number of contacts $Z$ in Fig.~\ref{fgr:viscosity} $(c)$ and $(d)$, for frictionless particles, the two protocols with different $\phi$'s collapse on each other despite having both different packing fractions and orientational properties. Similar to the disc and sphere cases~\cite{dong2020transition, ness2017oscillatory}, the low oscillation strains follow the steady shear curve, but unlike the isotropic case, only up to a $Z$ value of roughly 4 (the maximum value for discs), after which the viscosities are consistently lower than the corresponding SS case. For frictional particles, the viscosities coincide with SS up to roughly $Z=3$, after which small oscillatory strains differ both compared to frictional and frictionless SS, with a divergence close to $Z=4$, the value at which frictionless discs jam. Decomposing the rheological response as a function of the oscillatory viscous number $J'$, we find that the complex stress ratios $\mu^\ast(J')$ (see SI for a precise definition and data) for frictionless suspensions at low $\gamma_0$ differ between our two preparation protocols at large $J'(>0.1)$ values with non-directional suspensions having higher $\mu^*$ compared to the pre-sheared ones (see Fig. S2). For $\gamma_0>1$, the protocols yield the same curves. All of the curves seem to collapse at low $J'$ values irrespective of the strain amplitude.
Similarly, for frictional ellipses, we find statistically identical $\mu^*$ irrespective of $\gamma_0$ and all studied $J'$-values.

\noindent Like for the isotropic case, we get smaller $\mu^*$ in the low $J'$-regime at small $\gamma_0$-values for both frictional and frictionless ellipses \cite{dong2020oscillatory}. But, unlike disc case, $\mu^*$ does not collapse on the SS curve for large $\gamma_0$ values. We speculate that this is due to the slow reorientation of the ellipses upon shear-reversal and/or slower compaction/dilatancy upon increased shear rates compared to the disc case. \\

\noindent\emph{Summary -} We have shown that dense suspensions composed of elongated particles ($\alpha=3)$ get orientationally arrested at low oscillatory strains (below $\gamma_0 \simeq 0.3$) if frictionless. Adding friction or having large oscillatory strains help the system to escape these states. For these frictionless ellipses, this arrest results in two or more oscillatory shear jamming points at these strains. Starting from a random configuration frictionless ellipses (oscillatory) shear jams at a lower packing fraction than starting from an ordered pre-sheared sample. 
The former packing fraction does, however, coincide with the packing fraction where directional shear jamming appears for frictionless ellipses \cite{trulsson2021reverse}.
The viscosity of frictionless suspensions can be largely correlated with their average number of contacts, even though there are apparent differences between OS and SS close to shear jamming at low oscillatory strains. Surprisingly, the mechanical response close to (oscillatory shear) jamming is completely uncorrelated to a suspension orientational configuration.  
In contrast to frictionless particles as well as frictional discs, we \cite{dong2020oscillatory} find that small oscillation strains increase the number of contacts at shear jamming for frictional ellipses rather than lower it. If this is due to the orientationally disordered state or not remains to be investigated.  \\
As for tapping \cite{Yuan2021edwards}, oscillations are thought to equilibrate these athermal systems, leading to a unique equation of state of a ``thermalised'' athermal system (\emph{i.e.,}~the Edwards conjecture \cite{Baule2018edwards}). Hence, finding two packing fractions violates this conjecture for small oscillatory strains, as ergodicity seems not to hold while fulfilled for the frictional cases. Instead, the frictionless cases end up in a dynamically arrested state, similar to a glassy state, keeping its original nematic ordering and the average particle orientation indefinitely. These findings shed light on the importance of preparation for granular systems composed of elongated particles, \emph{e.g.}~such as rice and cereals, on their mechanical properties and their reproducibility.  \\
We want to thank Junhao Dong for his early technical support on this work. The simulations were performed on resources provided by the Swedish National Infrastructure for Computing (SNIC) at the center for scientific and technical computing at Lund University (LUNARC). 

\appendix
\setcounter{figure}{0}
\makeatletter 
\renewcommand{\thefigure}{S\@arabic\c@figure}
\makeatother

%\begin{Appendixes}
\appendix

\section{Schematic view of the two preparation protocols}
\begin{figure}[h]
\includegraphics[scale=0.75]{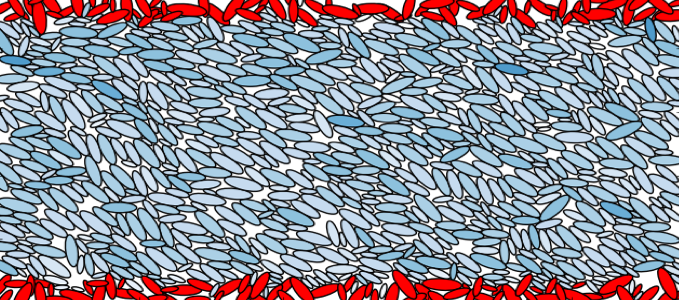}% Here is how to import EPS art
\hspace{4mm}
\includegraphics[scale=0.75]{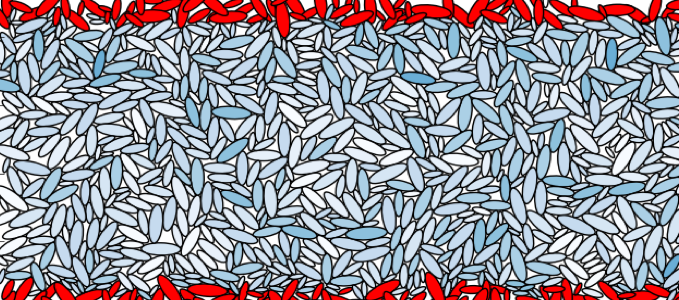}% Here is how to import EPS art
\caption{\label{sketch} Schematic view of the two preparation protocols: \emph{(up)} pre-sheared and \emph{(down)} random/non-directional.}   
\end{figure}

Fig.~\ref{sketch} shows a schematic view of our two preparation protocols: pre-sheared and random/non-directional.

\section{Complex stress ratio $\mu^*$, packing fraction $\phi$, number of contacts $Z$ and nematic order $S_2$ as a function of the viscous number}
\begin{figure}
\includegraphics[scale=0.6]{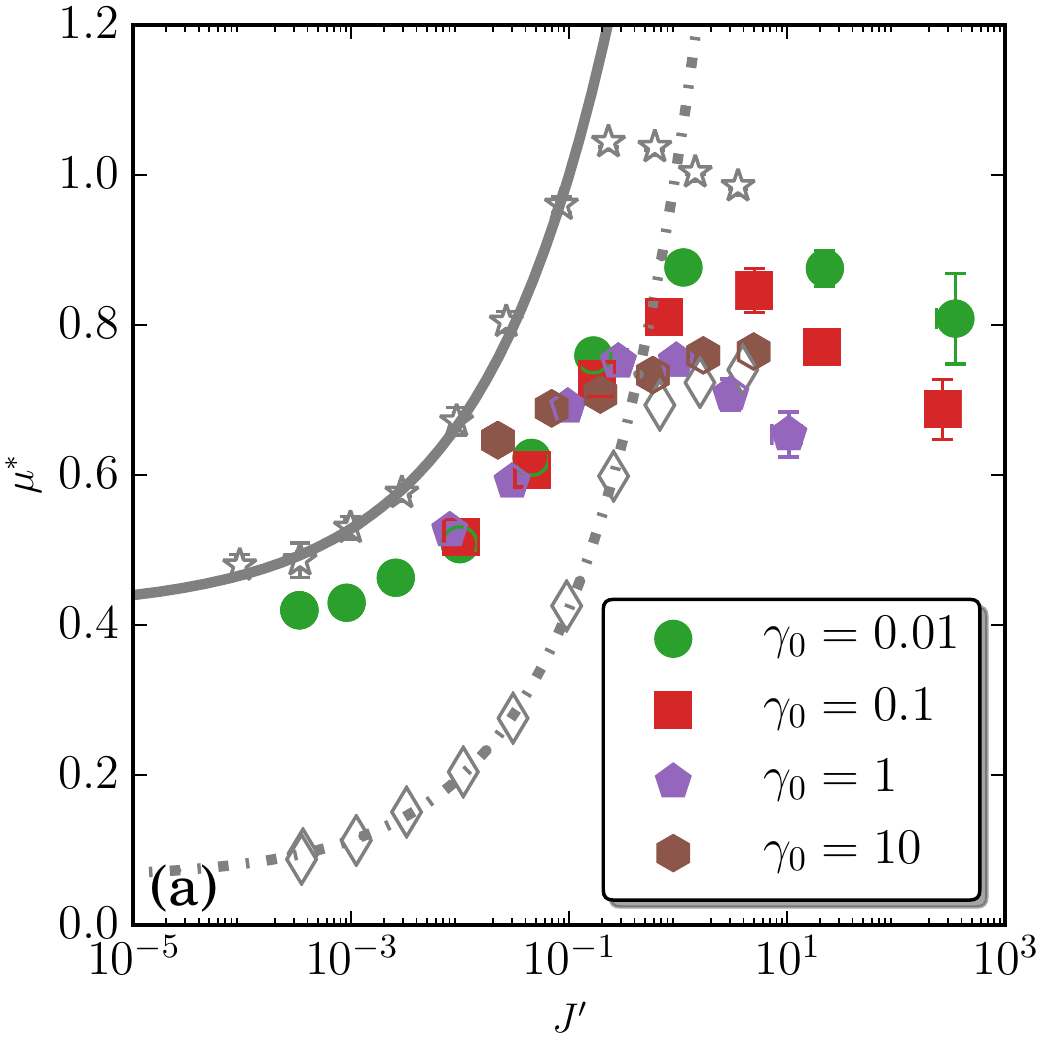}% Here is how to import EPS art

\includegraphics[scale=0.6]{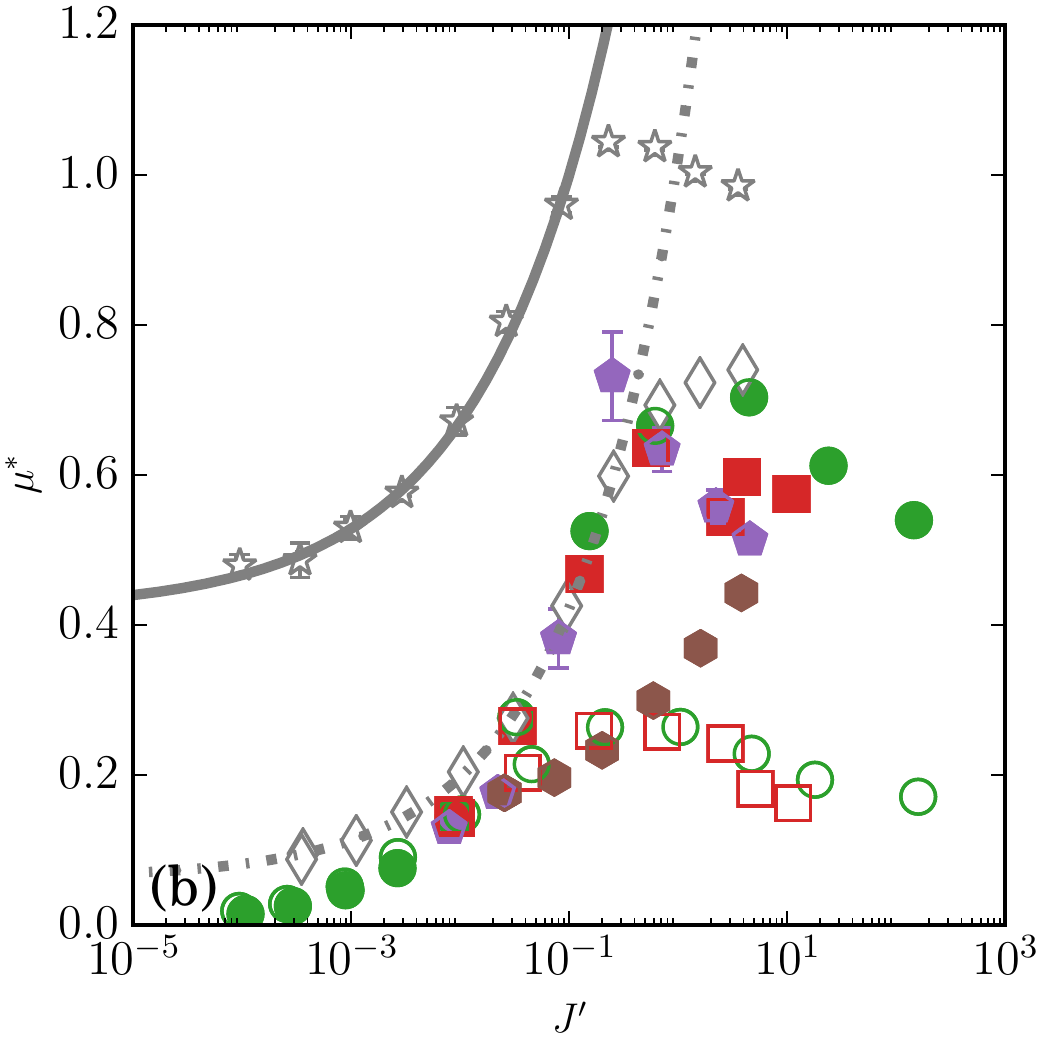}% Here is how to import EPS art
\caption{\label{complex} Complex friction coefficient $|\mu^\ast|$ for $(a)$ frictional and $(b)$ frictionless particles versus the viscous number $J'$ at various strain magnitudes $\gamma_0$ for pre-sheared (open symbols) and random (full symbols) preparations. 
The grey solid and dashed curves are the corresponding $\mu^*$ profiles $\mu^*=\mu^*_c+a_{\mu}J'^{n_{\mu}}$ under steady shear for frictional and frictionless suspensions, respectively. $a_{\mu}=1.34$ and $n_{\mu}=0.36$ for frictional and $0.96$ and $0.43$ for frictionless suspensions, accordingly. The critical friction coefficient  $\mu^*_c$ for frictional is $0.42$ and for frictionless is $0.06$. }   
\end{figure}

Following the same approach as \cite{dong2020transition}, we calculate shear rate averaged and strain averaged stress ratios $\mu'$ and $\mu''$, respectively as 

\begin{eqnarray}
\mu^{\prime}=\frac{\int_{0}^{2\pi/\omega} (\sigma/P)(t)\dot\gamma(t) \, dt}{\int_{0}^{2\pi/\omega} |\dot\gamma(t)| \, dt}
\end{eqnarray}
\begin{eqnarray}
\mu^{\prime\prime}=\frac{\int_{0}^{2\pi/\omega} (\sigma/P)(t)\gamma(t) \, dt}{\int_{0}^{2\pi/\omega} |\gamma(t)| \, dt}
\end{eqnarray}

where $\mu'$ is the viscous component and $\mu''$ is the elastic component
of the complex macroscopic friction coefficient $\mu^*$. The magnitude of $\mu^*$ is $|\mu^\ast|=\sqrt{\mu^{\prime2}+\mu^{\prime\prime2}}$. Other properties such as the packing fraction $\phi$, the number of contacts $Z$ (as previously shown in \cite{dong2020oscillatory}), and the nematic ordering $S_2$ of the particles are not very sensitive to the averaging process if they are shear-rate-averaged, strain-averaged or time-averaged. Here, we calculate the shear-rate-weighted average of these quantities as

\begin{eqnarray}
A_{|\dot\gamma|}=\frac{\int_{0}^{2\pi/\omega} A(t)|\dot\gamma(t)| \, dt}{\int_{0}^{2\pi/\omega} |\dot\gamma(t)| \, dt}
\end{eqnarray}

where $A$ is either $\phi$, $Z$, or $S_2$. %$\theta_{\bold e \cdot \hat{y}}$, or $S_2$.
Fig.~\ref{complex} illustrates complex stress ratios $|\mu^\ast|$ versus the viscous number $J'$ for pre-sheared (open symbols) and random (full symbols) preparations composed of $(a)$ frictional and $(b)$ frictionless particles.
The complex stress ratios $\mu^\ast(J')$ for frictionless suspensions at low $\gamma_0$ differ between our two preparation protocols at large $J'(>0.1)$ where pre-sheared suspensions have lower $\mu^*$ compared to the non-directional. For $\gamma_0>1$, the protocols yield the same curves. All of the $\mu^*$ curves seem to collapse at low $J'$ values regardless of $\gamma_0$. For frictional ellipses in the same manner, we find statistically identical $\mu^*$ for our both protocols at all $\gamma_0$ and $J'$-values.

\begin{figure}
\includegraphics[scale=0.6]{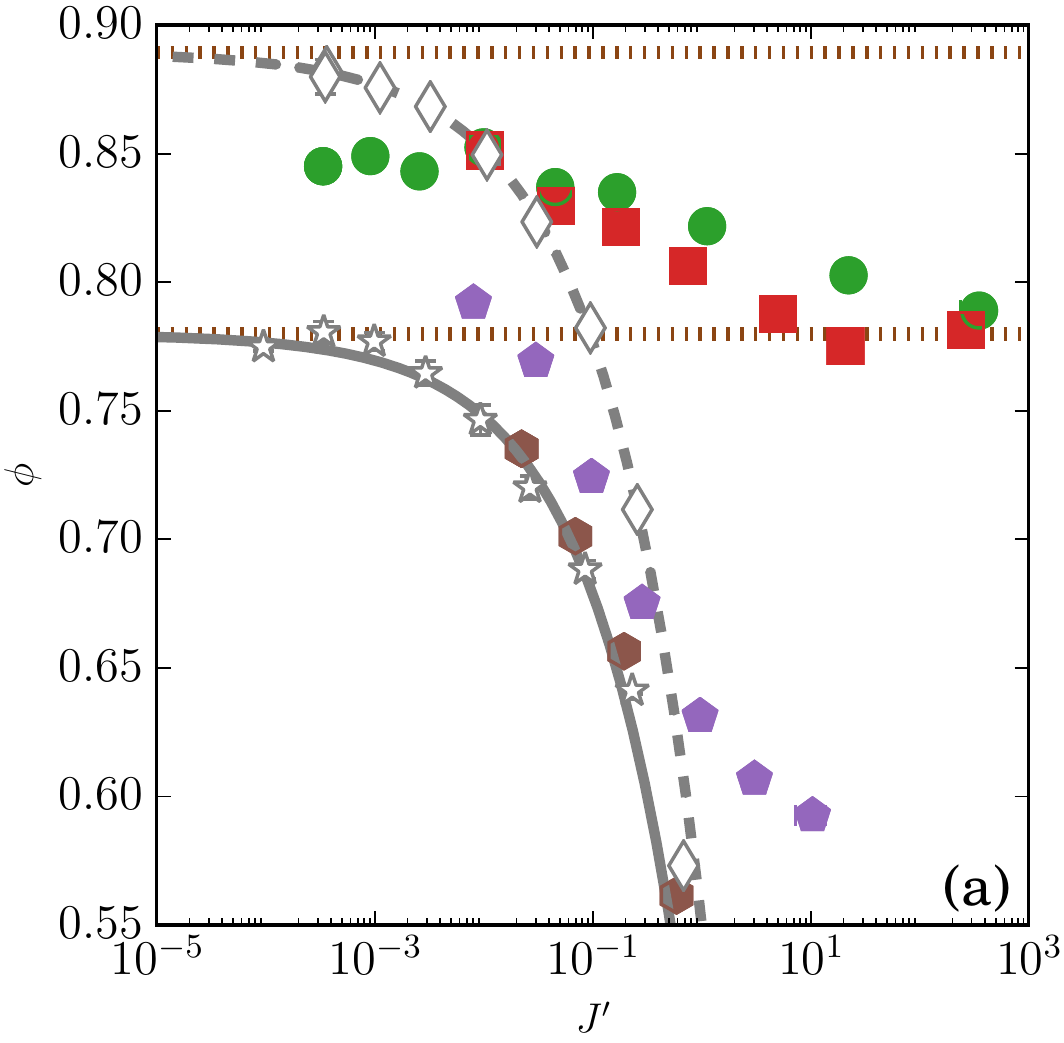}% Here is how to import EPS art

\includegraphics[scale=0.6]{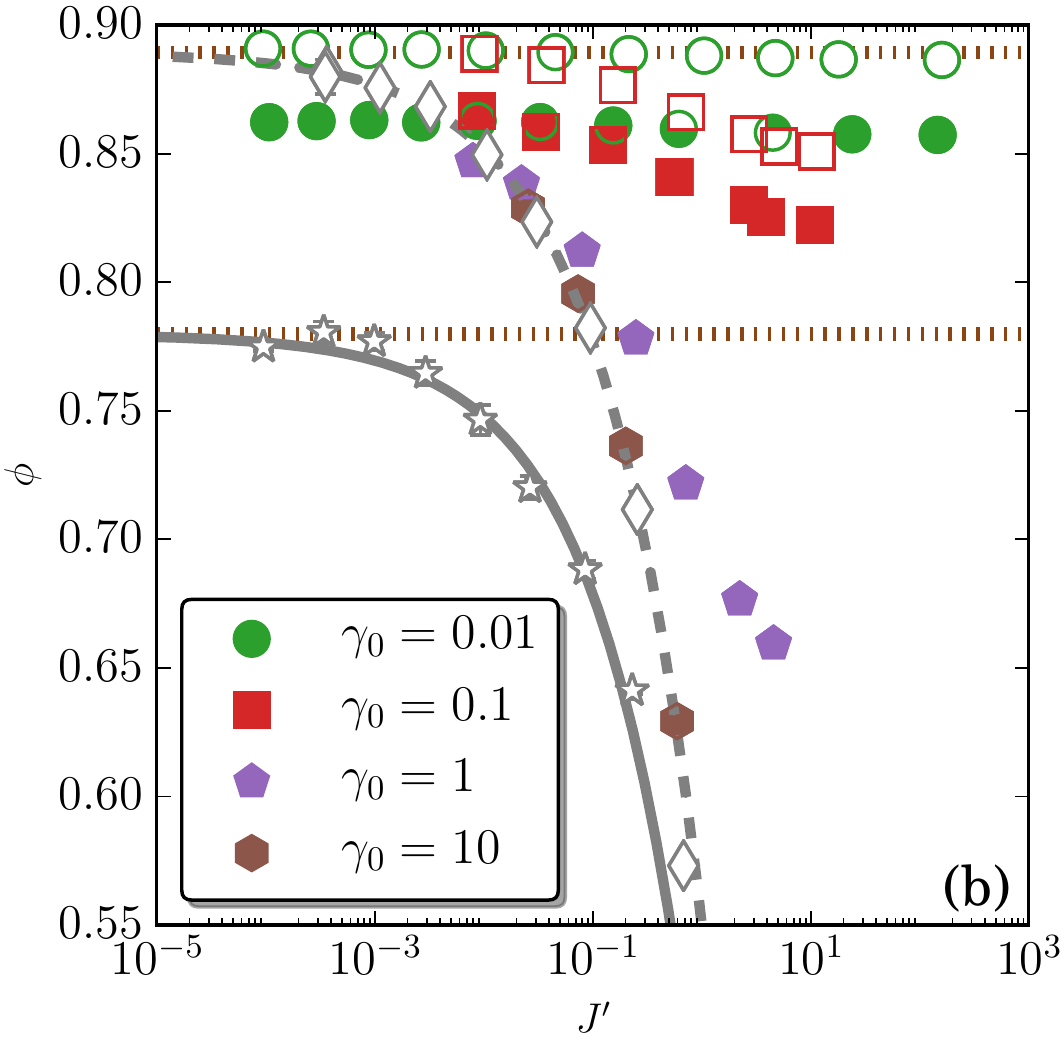}% Here is how to import EPS art
\caption{\label{phi}Packing fraction $\phi$ profiles versus viscous number $J'$ for $(a)$ frictional $(\mu_p=0.4)$ and $(b)$ frictionless $(\mu_p=0)$ suspensions. Data for pre-sheared (open symbols) and non-directional (full symbols) preparations have been illustrated. The grey solid and dashed lines are the corresponding steady shear $\phi$ curves $\phi=\phi_c-a_{\phi}J'^{1/n_{\phi}}$ for frictional and frictionless particles, respectively. $a_{\phi}$ and $n_{\phi}$ are $0.32$ and $2.01$, for frictional and $0.34$ and $2.1$, for frictionless cases, respectively. The golden horizontal lines show the steady shear jamming packing fraction for frictional $\phi_c^\text{f,SS}= 0.78$ and frictionless $\phi_c^\text{nf,SS}=0.89$ suspensions.}   
\end{figure}

To better analyse the two preparation protocols, packing fraction as a function of $J'$ \cite{dong2020oscillatory} is illustrated in Fig.~\ref{phi} for $(a)$ frictional ($\mu_p=0.4$) and $(b)$ frictionless ($\mu_p=0$) suspensions corresponding to our protocols: pre-sheared (open symbols) and non-directional (full symbols). For both frictional and frictionless ellipses $\phi$ depends on the oscillatory strain $\gamma_0$ and $J'$ where we find a collapse on the SS values at large $\gamma_0$ and in the frictional an increased $\phi$ at lower $\gamma_0$ at the same $J'$ values. We always get a unique curve in the frictional case irrespective of the directional ordering of the initial configurations. However, for frictionless ellipses, see Fig.~\ref{phi}$(b)$, we see that the two initial configurations render two different curves at low $\gamma_0$ ($\gamma_0\leq0.1$), with higher packing fractions starting from a pre-sheared configuration.

\begin{figure}
\includegraphics[scale=0.6]{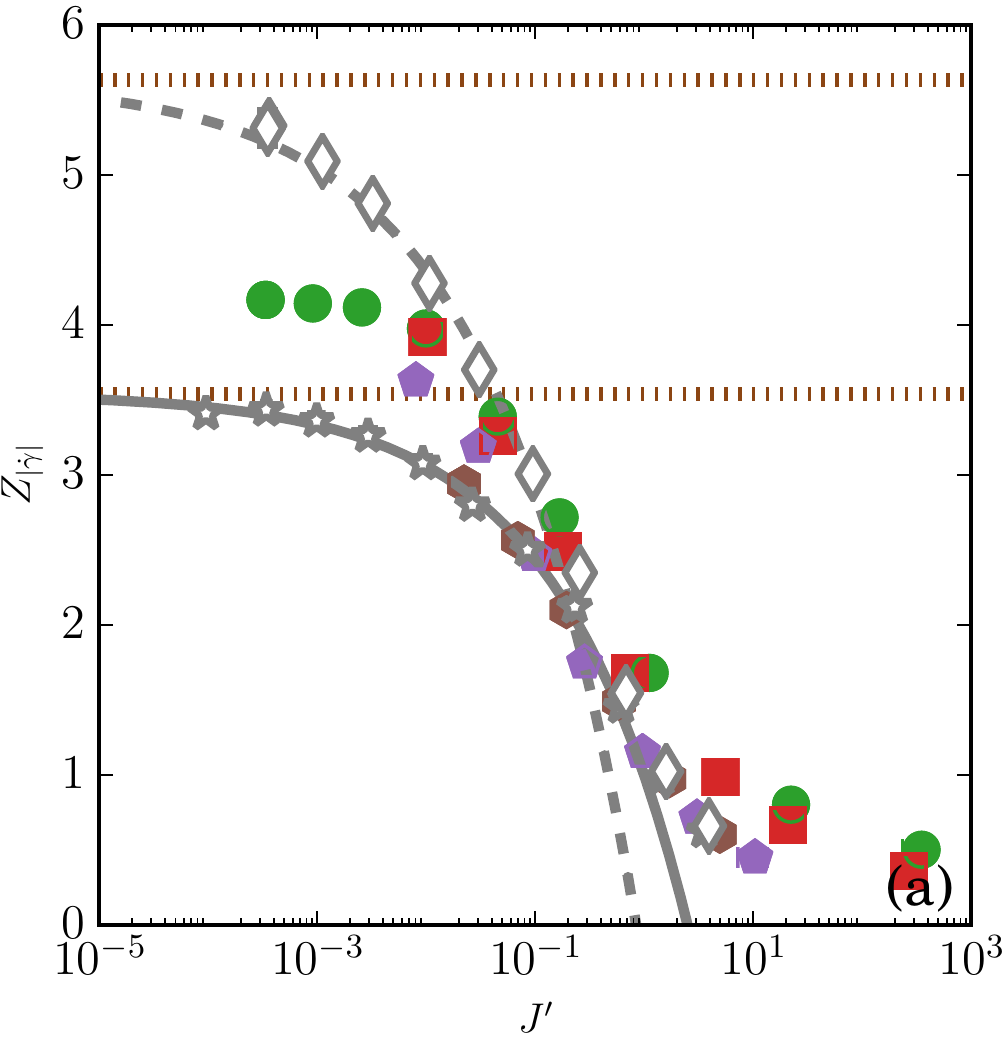}% Here is how to import EPS art

\includegraphics[scale=0.6]{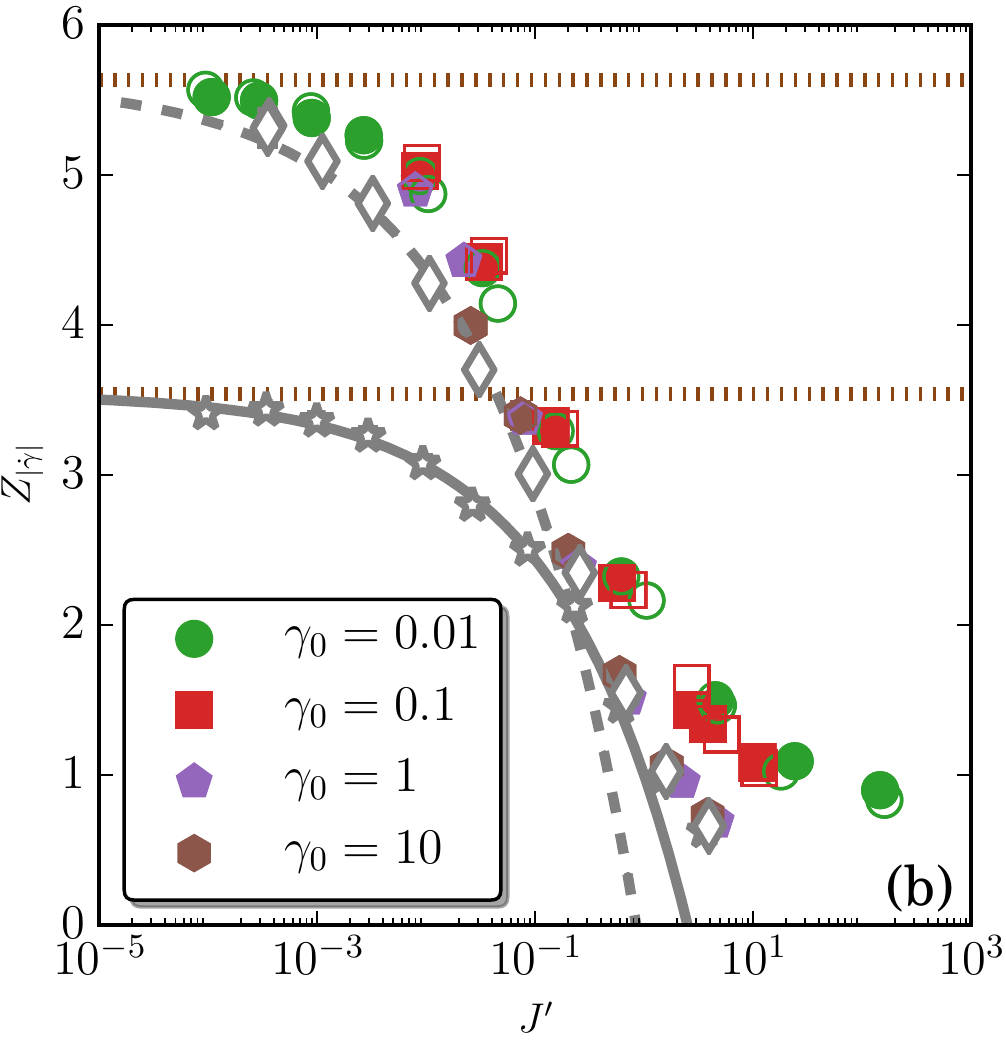}% Here is how to import EPS art
\caption{\label{Z}Number of contacts $Z$ versus viscous number $J'$ for $(a)$ frictional $(\mu_p=0.4)$ and $(b)$ frictionless $(\mu_p=0)$ suspensions. Data for pre-sheared (open symbols) and non-directional (full symbols) preparations have been illustrated. The gray solid and dashed lines are plots of the constitutive laws for the steady-shear cases. The constitutive laws are given as $Z=Z_c-a_ZZJ'^{1/n_{Z}}$, with $a_Z=2.54$ and
$n_Z=2.77$, for frictional and $a_Z=5.95$ and $n_Z=2.98$, for frictionless cases. The golden horizontal lines show the steady shear jamming number of contacts for frictional $Z_c^\text{f,SS}=3.54$ and frictionless $Z_c^\text{nf,SS}=5.63$ suspensions.}   
\end{figure}

Fig.~\ref{Z} shows the number of contacts for our two protocols as function of $J'$ for $(a)$ frictional and $(b)$ frictionless ellipses. We observe that the profiles of the two suspensions: pre-sheared and random, collapse on each other if we look at them from the perspective of the number of contacts.

\begin{figure}
\includegraphics[scale=0.6]{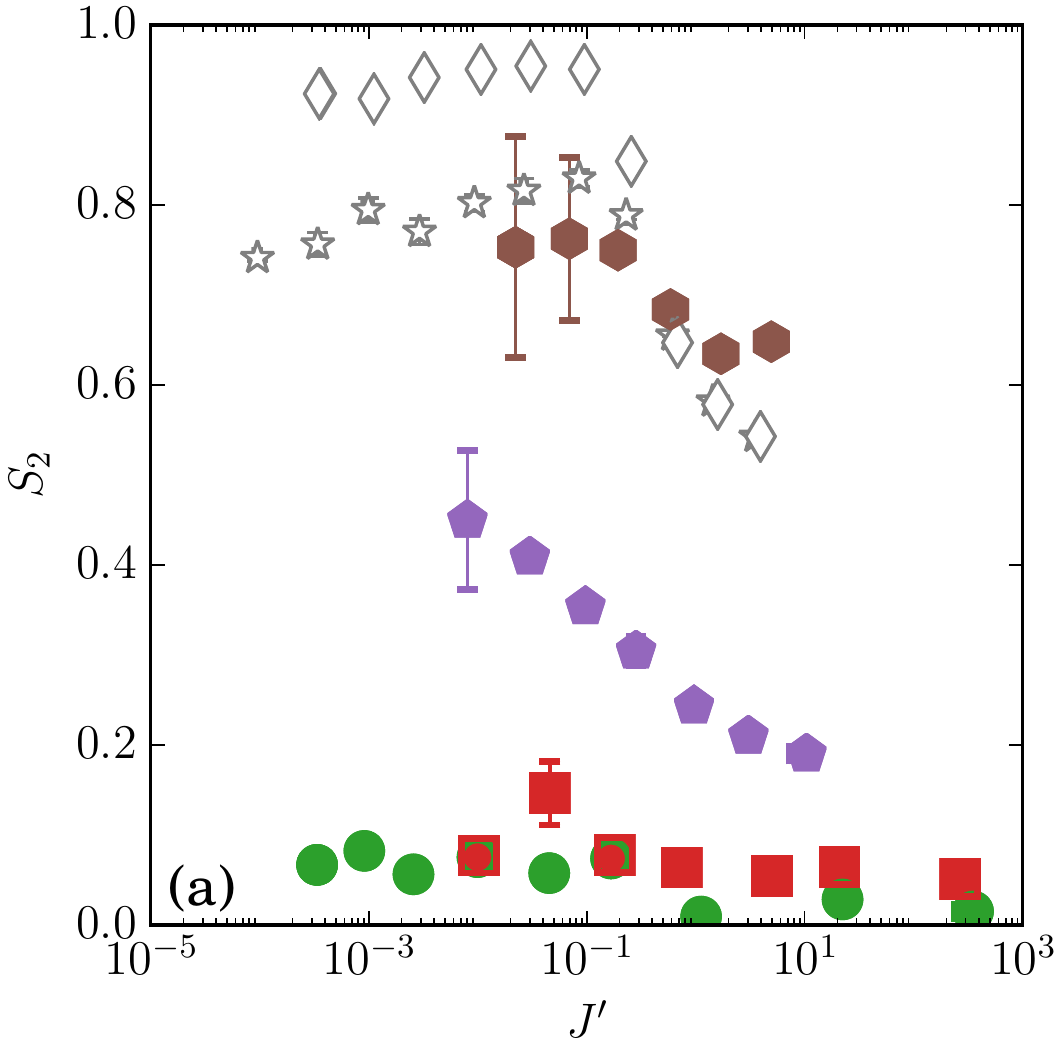}% Here is how to import EPS art

\includegraphics[scale=0.6]{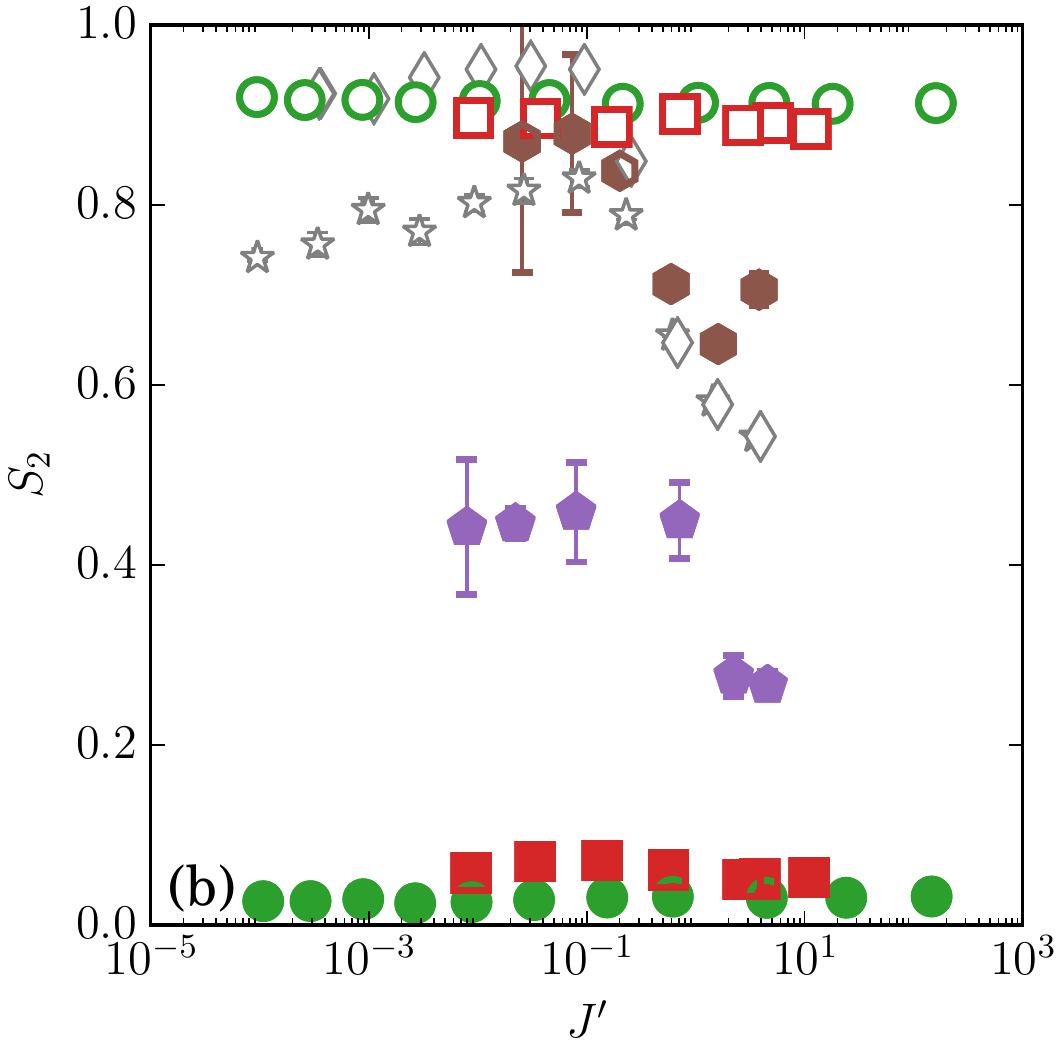}% Here is how to import EPS art
\caption{\label{S_2}Nematic order parameter $S_2$ versus viscous number $J'$ for $(a)$ frictional $(\mu_p=0.4)$ and $(b)$ frictionless $(\mu_p=0)$ suspensions. Data for pre-sheared (open symbols) and non-directional (full symbols) preparations have been illustrated. The grey stars and diamonds are the corresponding SS values for the frictional and the frictionless suspensions, respectively. The legends are the same as in Fig.~\ref{Z}.}   
\end{figure}

Fig.~\ref{S_2} shows the nematic ordering $S_2$ for our two protocols (presheared and random) as function of $J'$ for $(a)$ frictional and $(b)$ frictionless ellipses. We observe that in the frictionless case (Fig.~\ref{S_2}(b)) at low $\gamma_0$ the nematic order parameters of the pre-sheared and the random are different from each other (compare open and full symbols). Generally the nematic order is an increasing function of the oscillatory strain approaching toward its SS value at large $\gamma_0$. 

\section{Relaxation parameter of the nematic order $\kappa_{S_2}^{-1}$ as function of $\mu_p$}

\begin{figure}
\centering
\includegraphics[scale=0.6]{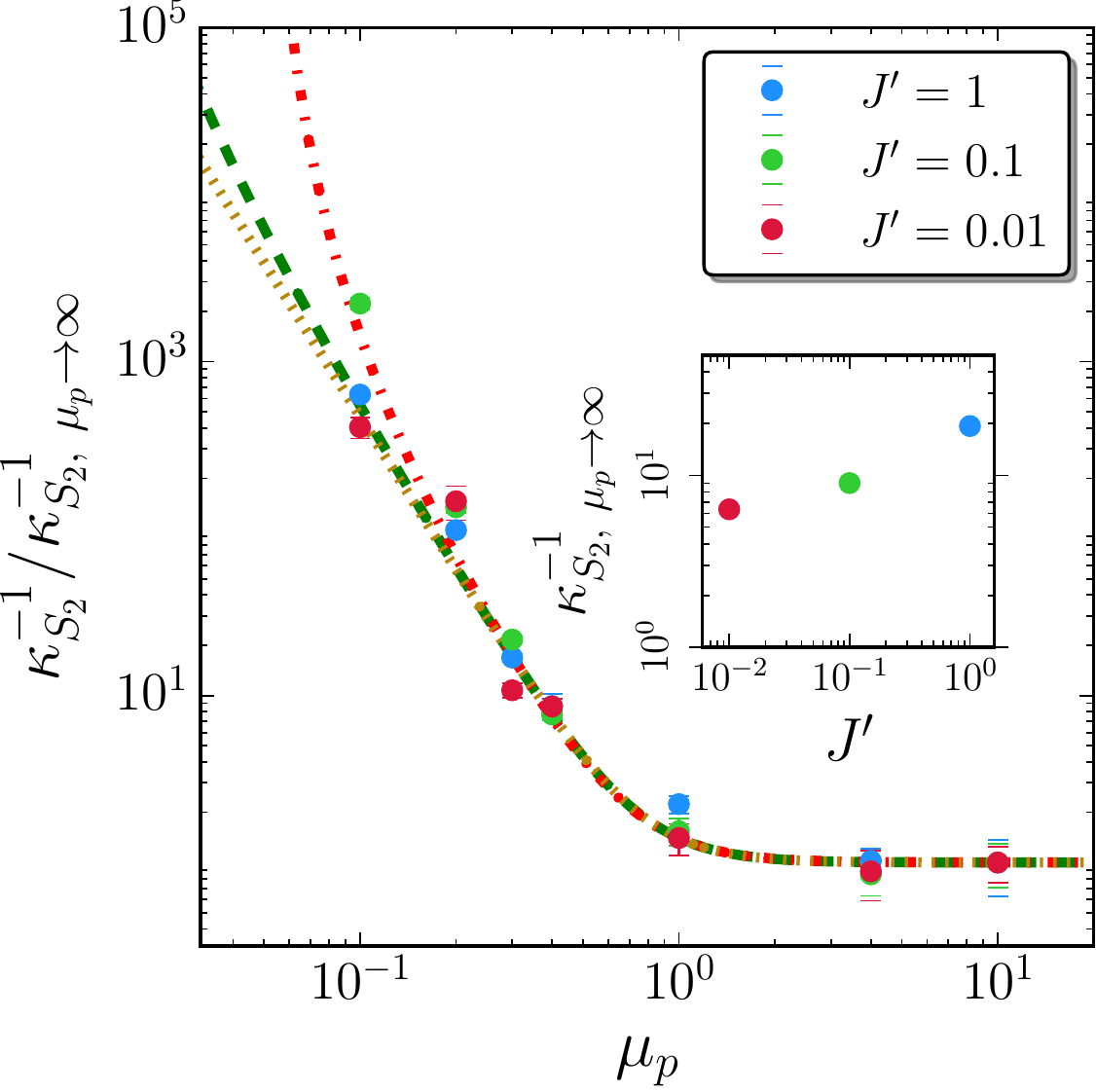}% Here is how to import EPS art
\caption{\label{relaxmu} Rescaled relaxation parameter of the nematic order $\kappa_{S_2}^{-1}$ for the pre-sheared configuration as function of $\mu_p$ at $\gamma_0=0.1$ for various $J'$. Each $J'$-curve has been normalised by its corresponding value at $\mu_p \to \infty$. Dashed lines indicate best fits of the data with the function $\kappa_{S_2}^{-1}\sim a_{S_2}(\mu_p-\mu_{p,c})^{-\beta}+ c$  with different coefficients, exponents and critical friction coefficient $\mu_{p,c}$: golden: $a_{S_2}=0.40$, $\beta=3.06$ and $\mu_{p,c}=0$, green: $a_{S_2}=0.39$, $\beta=3.01$ and $\mu_{p,c}=0.01$ and red: $a_{S_2}=0.34$, $\beta=2.78$ and $\mu_{p,c}=0.05$. For all the fittings $c$ is $1$.}
\end{figure}

Relaxation parameter $\kappa_{S_2}^{-1}$ as function of $\mu_p$ has been plotted in Fig.~\ref{relaxmu} at $\gamma_0=0.1$ for various $J'$. We have used the function $\kappa_{S_2}^{-1}\sim a_{S_2}(\mu_p-\mu_{p,c})^{-\beta}+ c$ to fit the data. According to Fig.~\ref{relaxmu}, due to the sharp divergence of the data, it is hard to tell if the divergence occurs at a finite $\mu_{p,c}$ or $\mu_{p,c}=0$. In the figure, we have shown a few possible fittings using the same function form but with different fitting parameters and divergence $\mu_{p,c}$ that all seem to explain our data well. 

\section{Relaxation parameter of the nematic order $\kappa_{S_2}^{-1}$ as function of oscillatory strain $\gamma_0$ (frictionless)}

\begin{figure}
\includegraphics[scale=0.6]{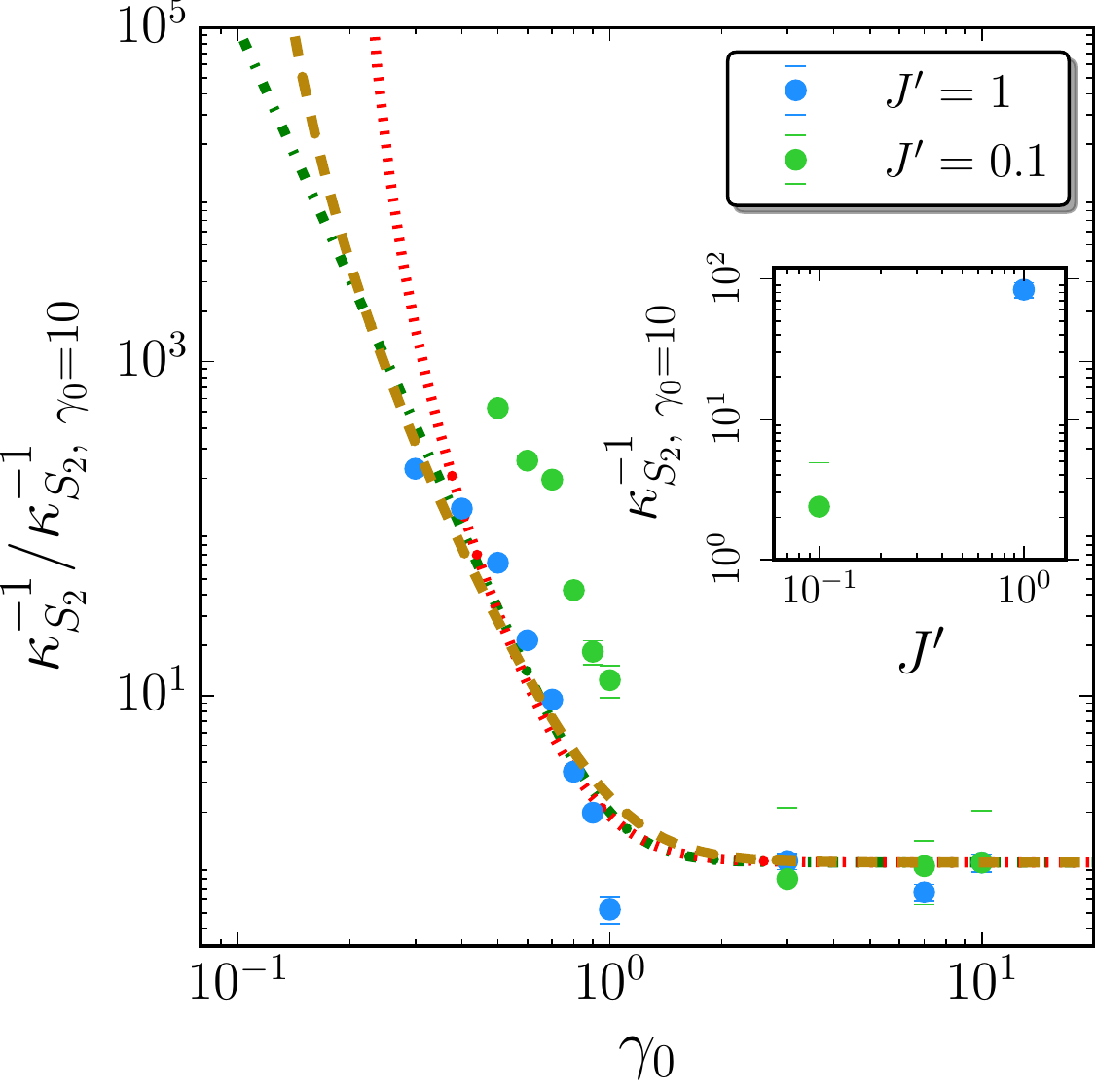}% Here is how to import EPS art
\caption{\label{relaxgamma} Rescaled relaxation parameter of the nematic order $\kappa_{S_2}^{-1}$ for the pre-sheared configuration as function of $\gamma_0$ at $\mu_p=0$ for various $J'$. Each $J'$-curve has been normalised by its corresponding value at $\gamma_0=10$. Dashed lines indicate best fits of the data with the function $\kappa_{S_2}^{-1}\sim a_{S_2}(\gamma_0-\gamma_{0,c})^{-\beta}$ + c with different coefficients, exponents and critical oscillatory strain $\gamma_{0,c}$: golden: $a_{S_2}=1$, $\beta=3.59$ and $\gamma_{0,c}=0.1$, green: $a_{S_2}=1$, $\beta=5$ and $\gamma_{0,c}=0$ and red: $a_{S_2}=0.40$, $\beta=3.60$ and $\gamma_{0,c}=0.2$. For all the fittings $C$ is $1$.}
\end{figure}

Relaxation parameter $\kappa_{S_2}^{-1}$ as function of $\gamma_0$ at $\mu_p=0$ has been plotted in Fig.~\ref{relaxgamma} for various $J'$. We have used the function $\kappa_{S_2}^{-1}\sim a_{S_2}(\gamma_0-\gamma_{0,c})^{-\beta}+ c$ to fit the data. According to Fig.~\ref{relaxgamma}, due to the sharp divergence of the data, it is hard to tell the exact divergence $\gamma_{0,c}$. In the figure, we have shown a few possible fittings using the same function form but with different fitting parameters and divergence $\gamma_{0,c}$.

\section{Pre-sheared and randomly ordered frictionless suspensions at low $\gamma_0$ (\emph{i.e.} $\gamma_0\leq0.1$)}

\begin{figure}
\includegraphics[scale=0.385]{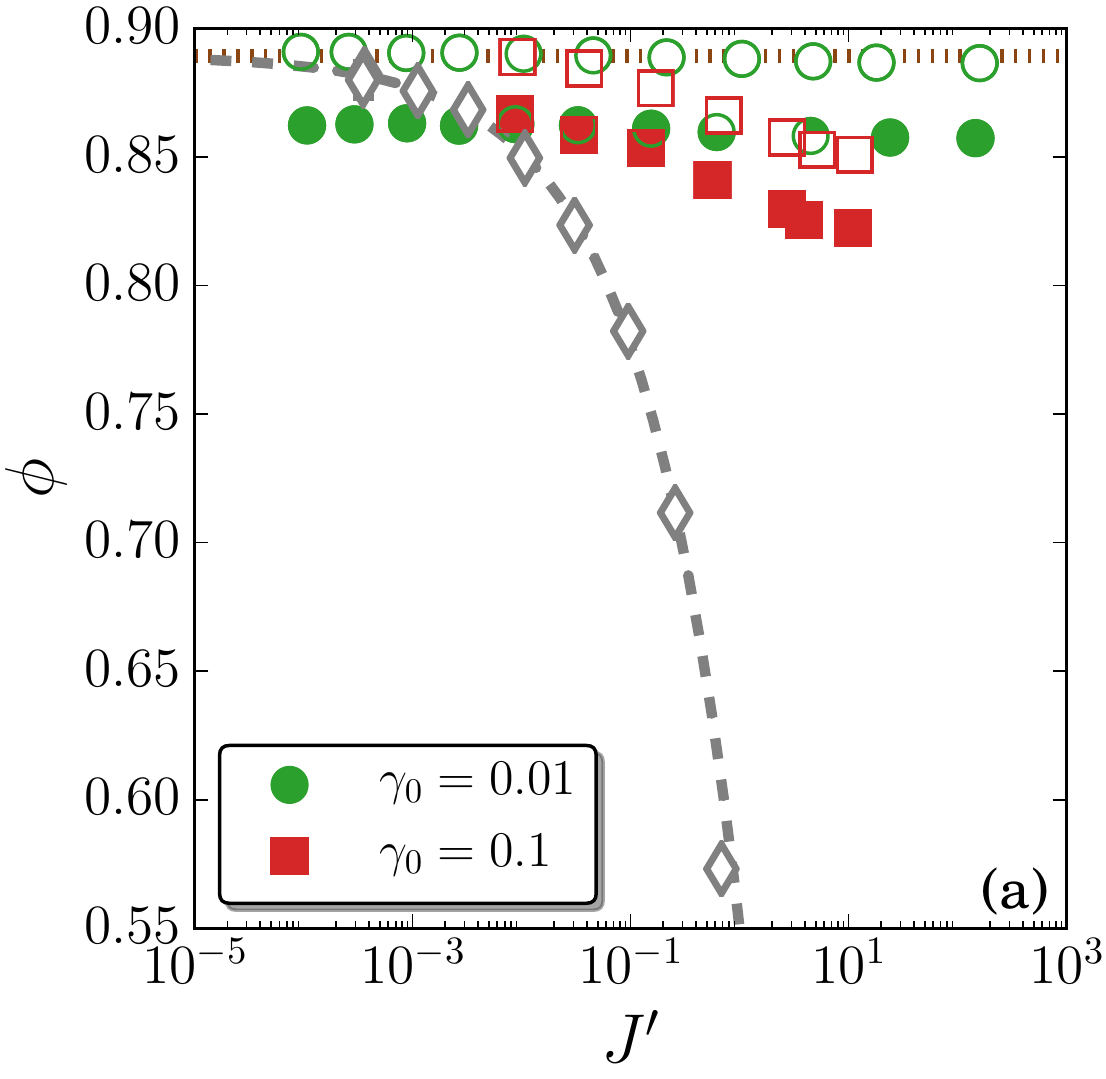}% Here is how to import EPS art
\includegraphics[scale=0.385]{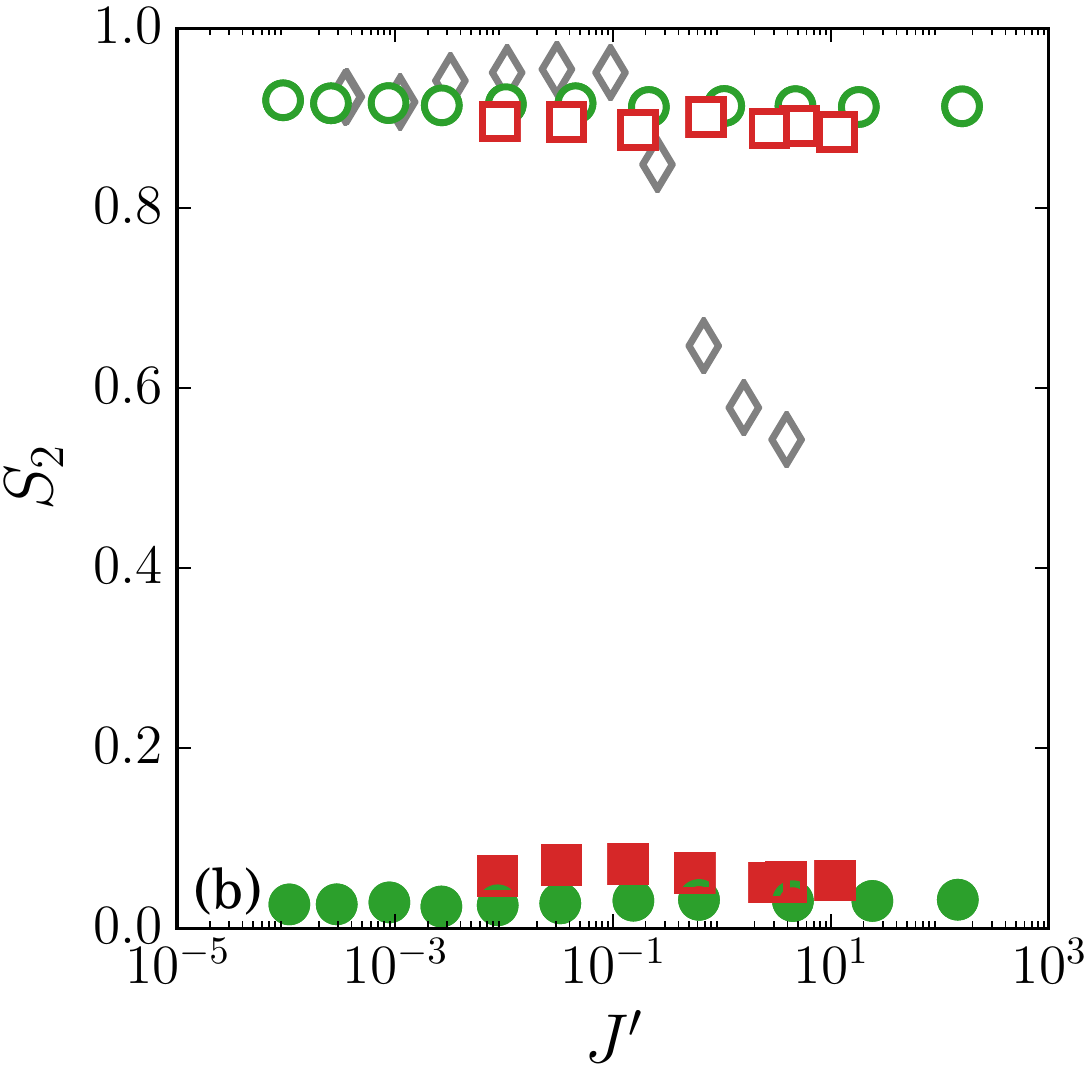}% Here is how to import EPS art

\includegraphics[scale=0.385]{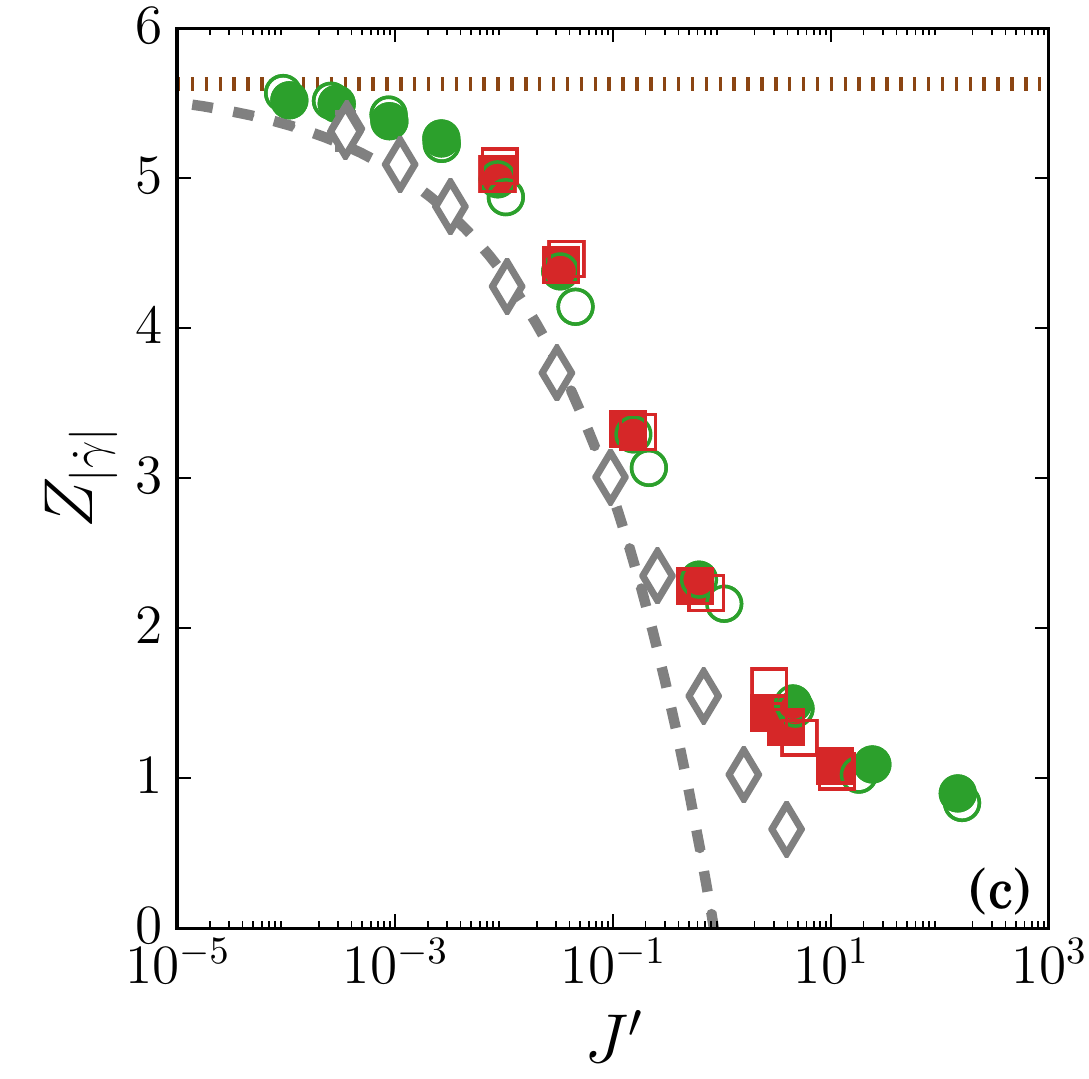}% Here is how to import EPS art
\includegraphics[scale=0.385]{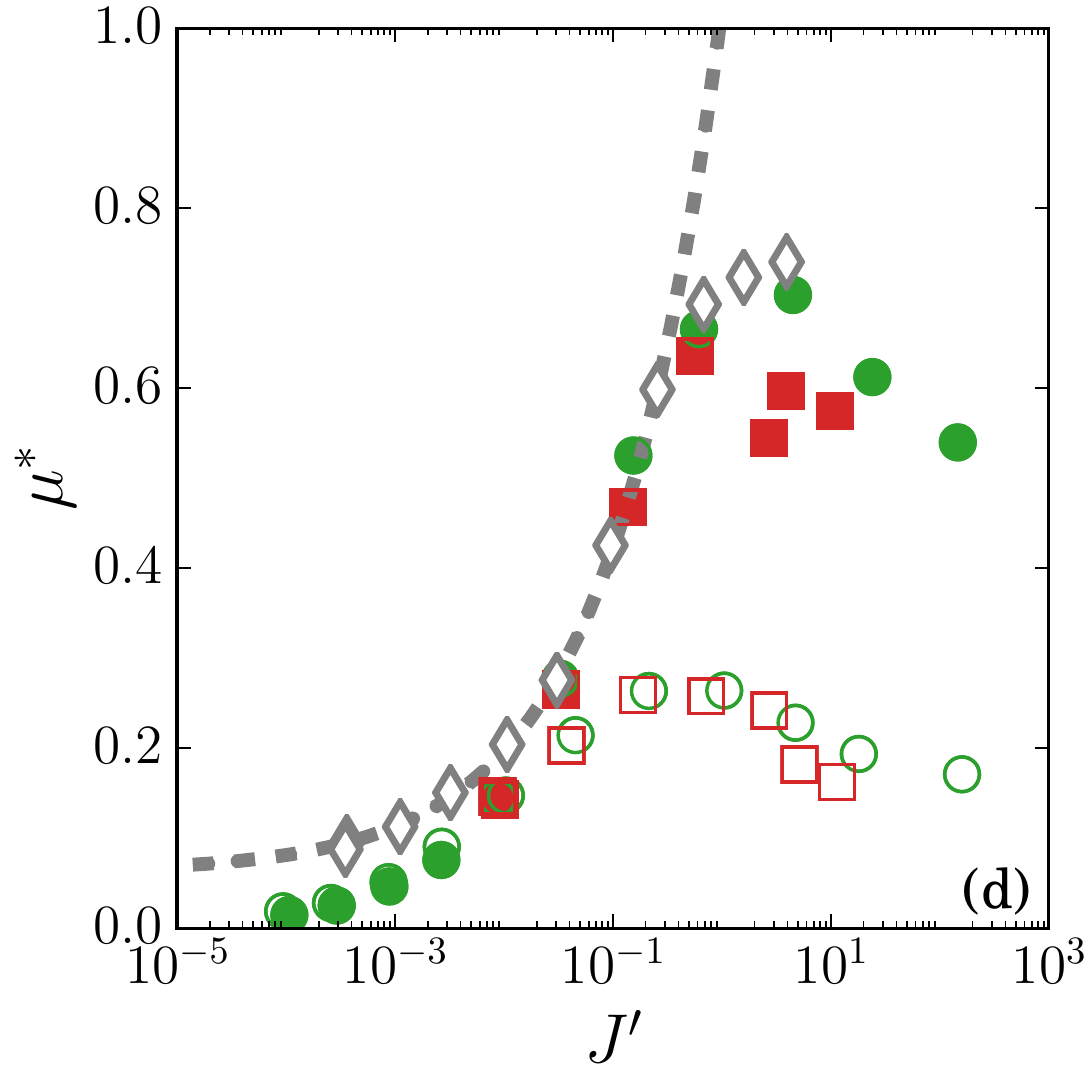}% Here is how to import EPS art
\caption{\label{Protocol} $(a)$ $\phi$, $(b)$ $S_2$, $(c)$ $Z_{|\dot\gamma|}$ and $(d)$ $|\mu^*|$ of the frictionless pre-sheared (open symbols) and random (full symbols) suspensions at low oscillatory strains versus viscous number $J'$. Grey diamonds show the corresponding data at steady shear. Dashed lines are fitted curves at steady shear.}   
\end{figure}

To better illustrate the role of the starting configuration at small oscillatory strains, we show $\phi$, $S_2$, $Z$, and $\mu^\ast$ as a function of $J'$ for the two protocols at our two lowest $\gamma_0$. Fig.~\ref{Protocol} shows that the initial orientation affects $\phi$, $S_2$ and $\mu^\ast$ but not $Z$.

\section{Forward/backward stress responses of the pre-sheared preparations}

As shown in Fig. 4 in the main text, at oscillatory strains less than or equal to $0.1$, we observe asymmetry in stress responses of the pre-sheared frictional configurations while being sheared along the pre-sheared direction compared to that opposite to it. However, the stress response for frictionless pre-sheared suspension is less asymmetric and within or close to our noise level.

\begin{figure*}
\includegraphics[scale=0.8]{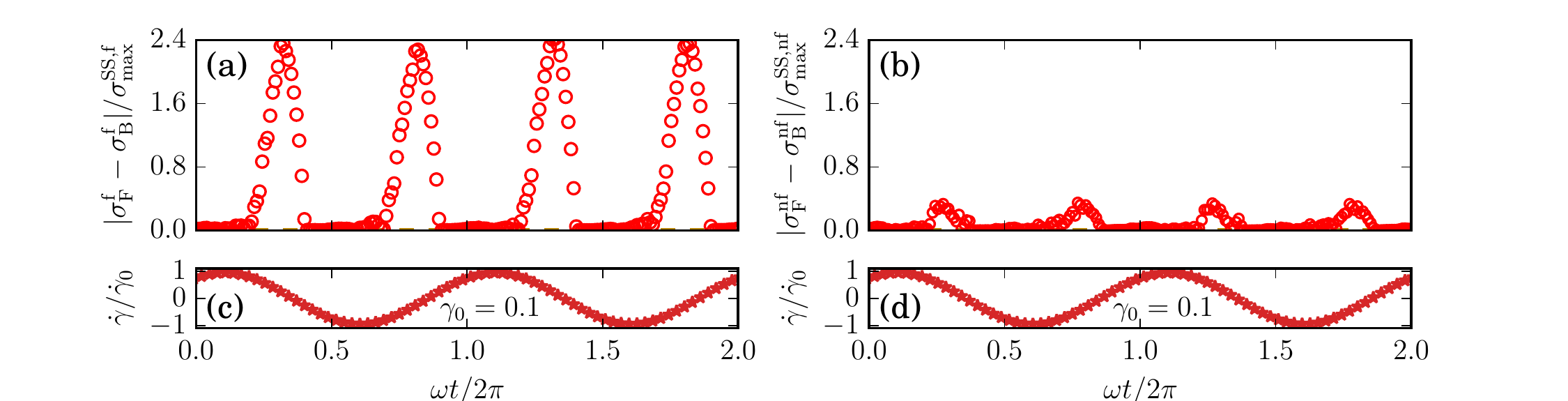}% Here is how to import EPS art
\caption{\label{AsymInit} Normalized difference between forward and backward shear stress of the pre-sheared at early stage of the stress evolution after $\phi$ has relaxed. $(a)$ belongs to the frictional and $(b)$ belongs to the frictionless configurations at $\gamma_0=0.1$ and $J'\simeq 0.1$. Subscripts $\text{F}$ and $\text{B}$ stand for 'Forward' and 'Backward', respectively. Figures $(c)$ and $(d)$ show the corresponding rescaled shear rate to $(a)$ and $(b)$, respectively.}
\end{figure*}

\begin{figure*}
\includegraphics[scale=0.8]{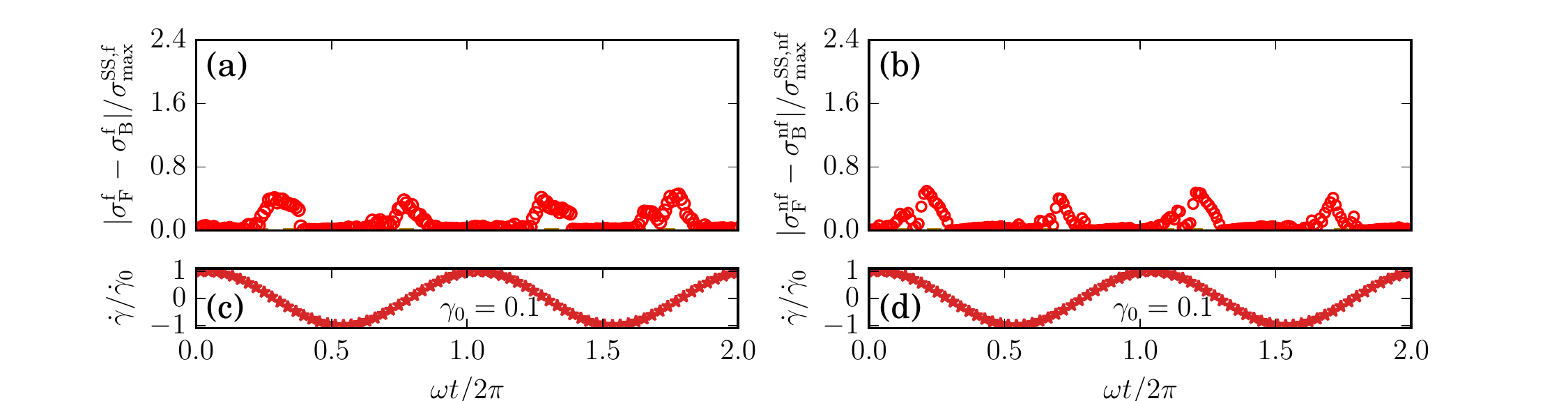}% Here is how to import EPS art
\caption{\label{AsymFinal} Re-scaled absolute discrepancy between forward and backward shear stress response after complete stress relaxation of the pre-sheared configuration. $(a)$ belongs to the frictional and $(b)$ belongs to the frictionless configurations at $\gamma_0=0.1$ and $J'=0.1$. Subscripts $\text{F}$ and $\text{B}$ stand for 'Forward' and 'Backward', respectively. Figures $(c)$ and $(d)$ show the corresponding rescaled shear rate to $(a)$ and $(b)$, respectively.}
\end{figure*}

\begin{figure*}
\includegraphics[scale=0.8]{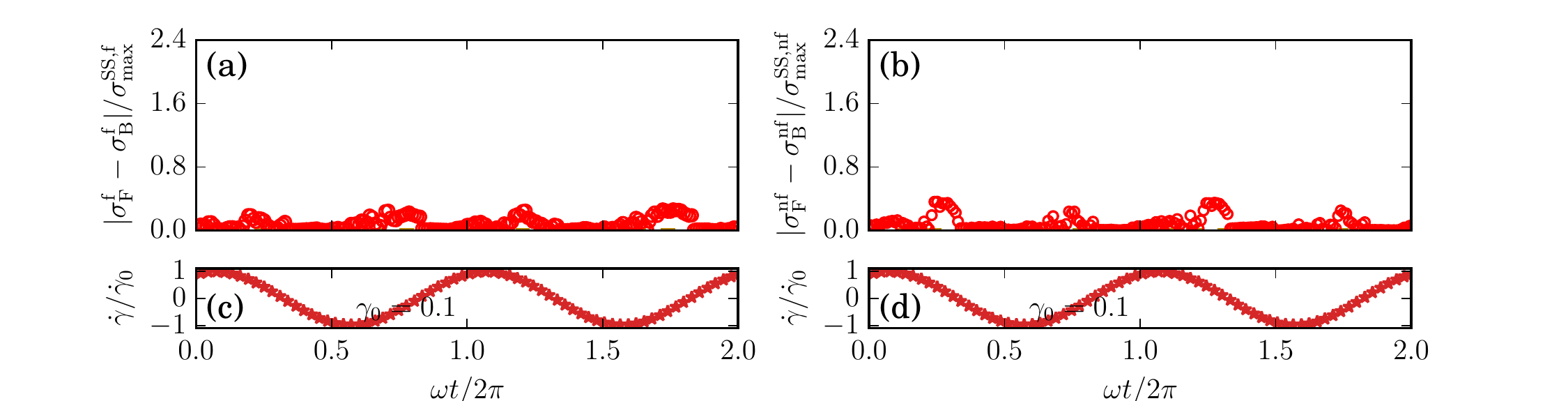}% Here is how to import EPS art
\caption{\label{AsymRan} Normalized difference between forward and backward stress response for the random configurations. $(a)$ belongs to the frictional and $(b)$ belongs to the frictionless configurations at $\gamma_0=0.1$ and $J'=0.1$. Subscripts $\text{F}$ and $\text{B}$ stand for 'Forward' and 'Backward', respectively. Figures $(c)$ and $(d)$ show the corresponding rescaled shear rate to $(a)$ and $(b)$, respectively.}   
\end{figure*}

To illustrate the effect more clearly, we plot the absolute difference between forward and backward stress responses in each period for three different cases: at an early stage of pre-sheared, late stage of pre-sheared, and of random/non-directional configuration. Fig.~\ref{AsymInit} shows that the forward and backward stress response is asymmetric at early cycles for frictional ellipses, but only mildly for frictionless, starting from a pre-sheared suspension. Stress evolutions shown are from after $\phi$ has relaxed (which relaxes a magnitude faster).
After some relaxation, the stress response is much less asymmetric, and the frictional and the frictionless curves reassemble each other, see Fig.~\ref{AsymFinal}. Notice that there persists a small asymmetry, but this most likely is a finite size effect, as starting from random initial configurations renders roughly the same asymmetry, see Fig.~\ref{AsymRan}.

\section{Time-evolutions of $\theta_{\bold e \cdot \hat{y}}$, $S_2$ and $\phi$}

\begin{figure*}[th!]

      \includegraphics[scale=0.62]{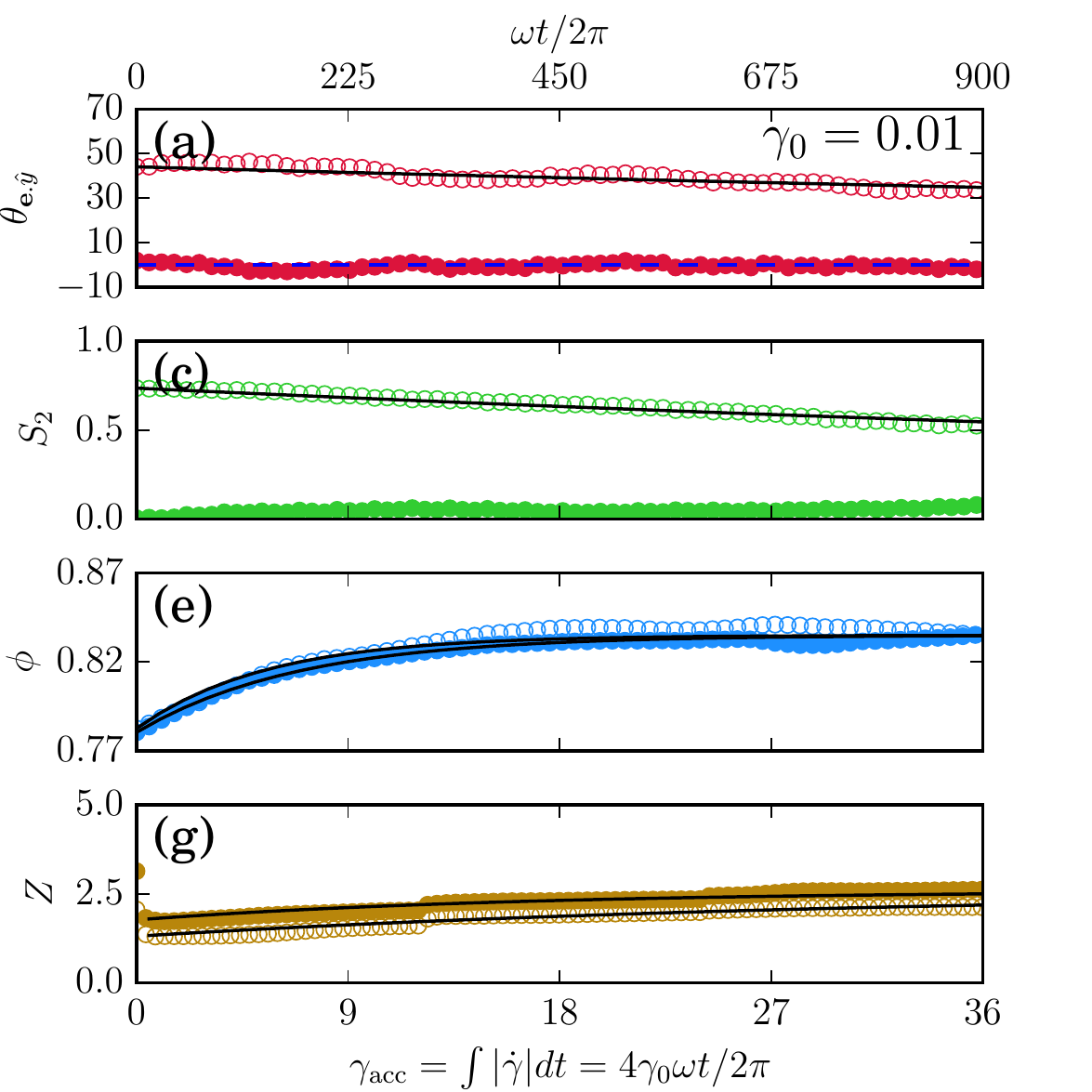}% Here is how to import EPS art
      \includegraphics[scale=0.62]{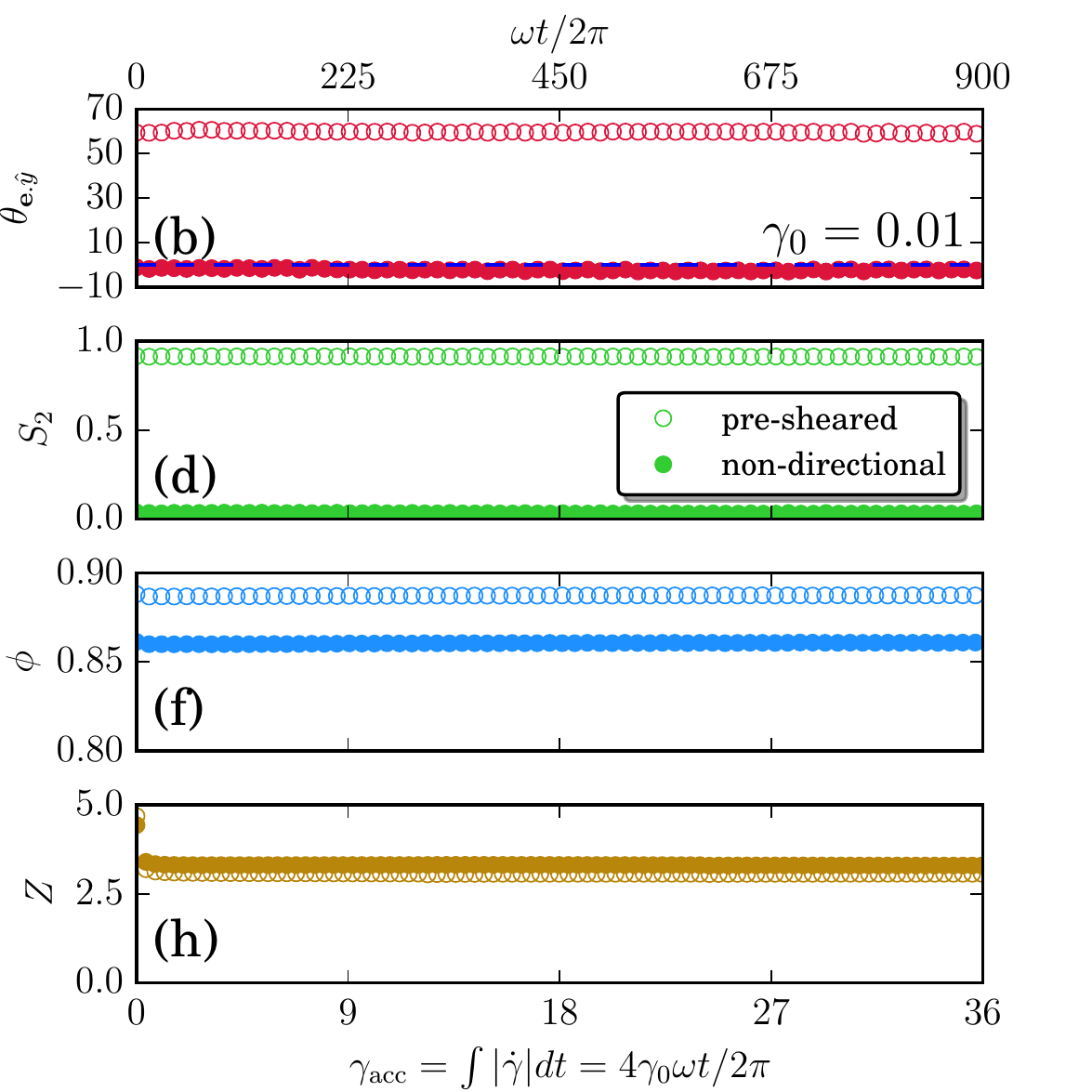}% Here is how to import EPS art

      \includegraphics[scale=0.62]{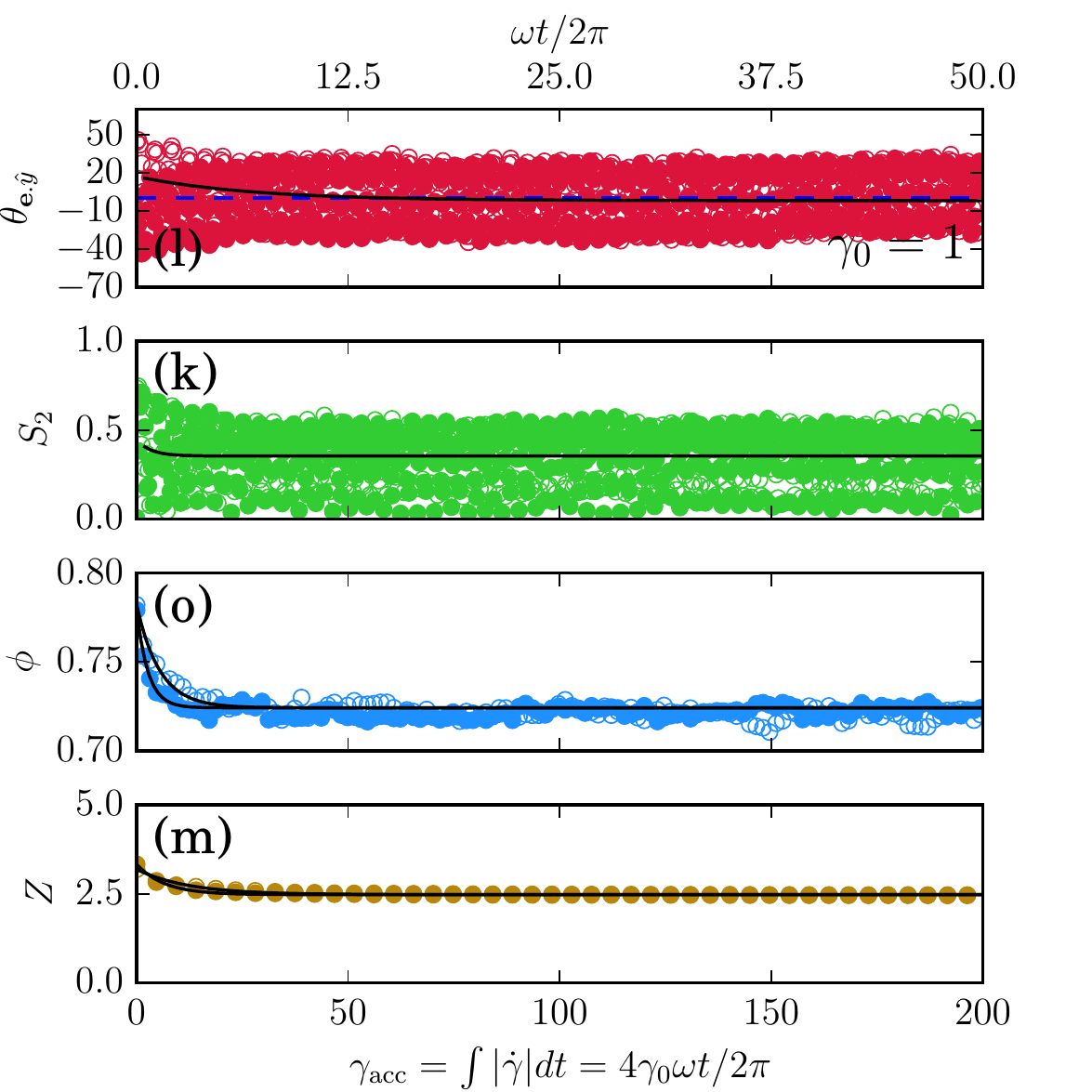}% Here is how to import EPS art
      \includegraphics[scale=0.62]{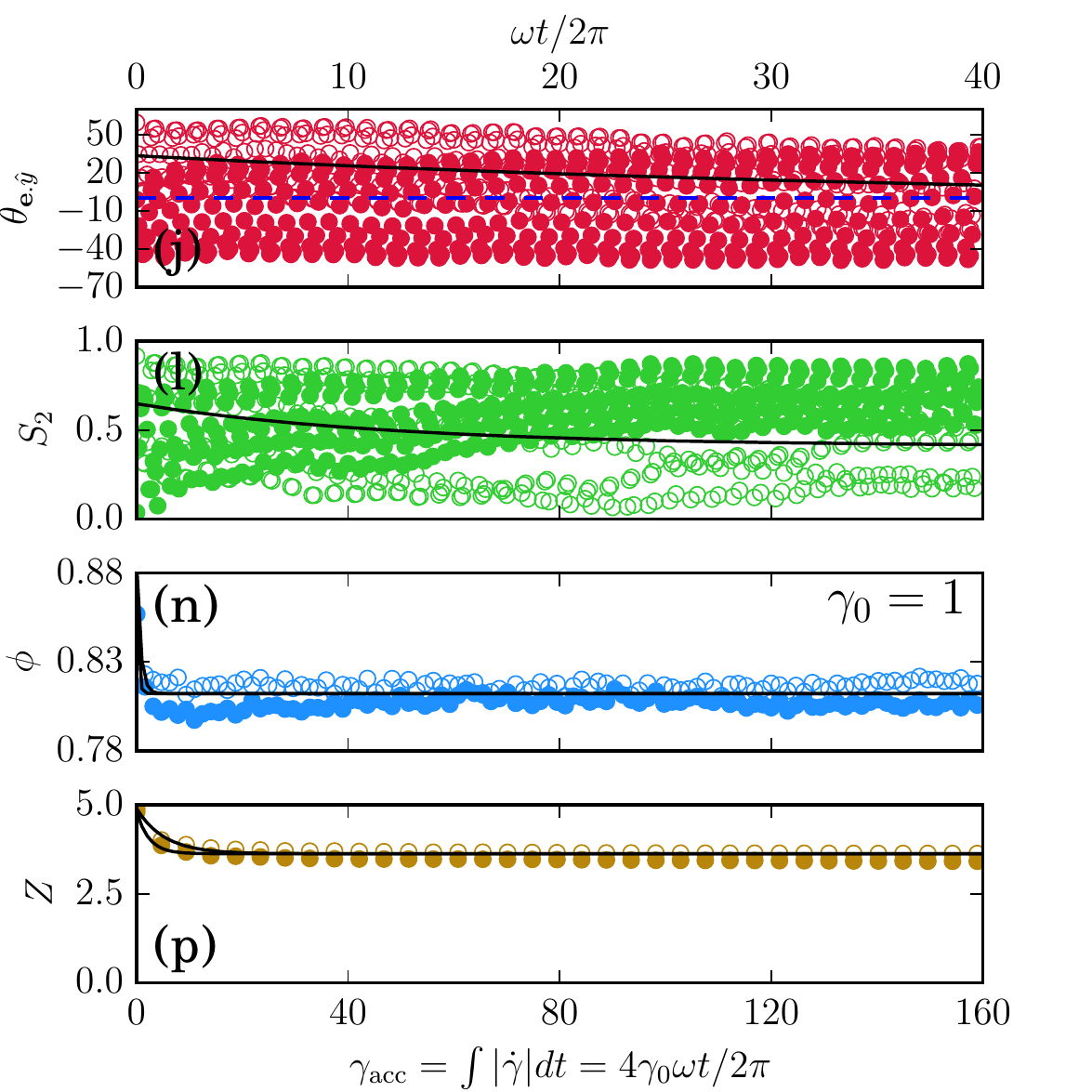}
      
    \includegraphics[scale=0.62]{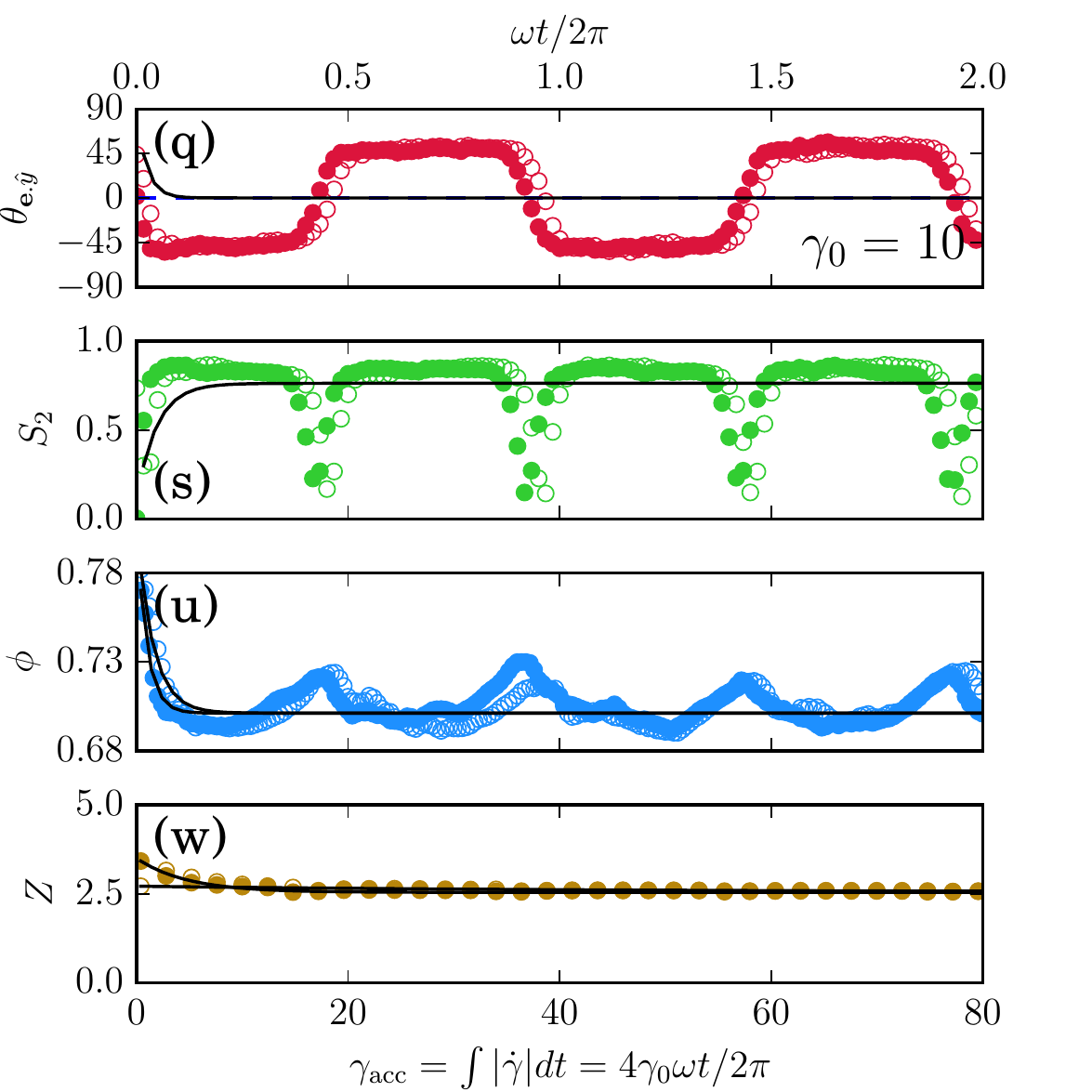}% Here is how to import EPS art
    \includegraphics[scale=0.62]{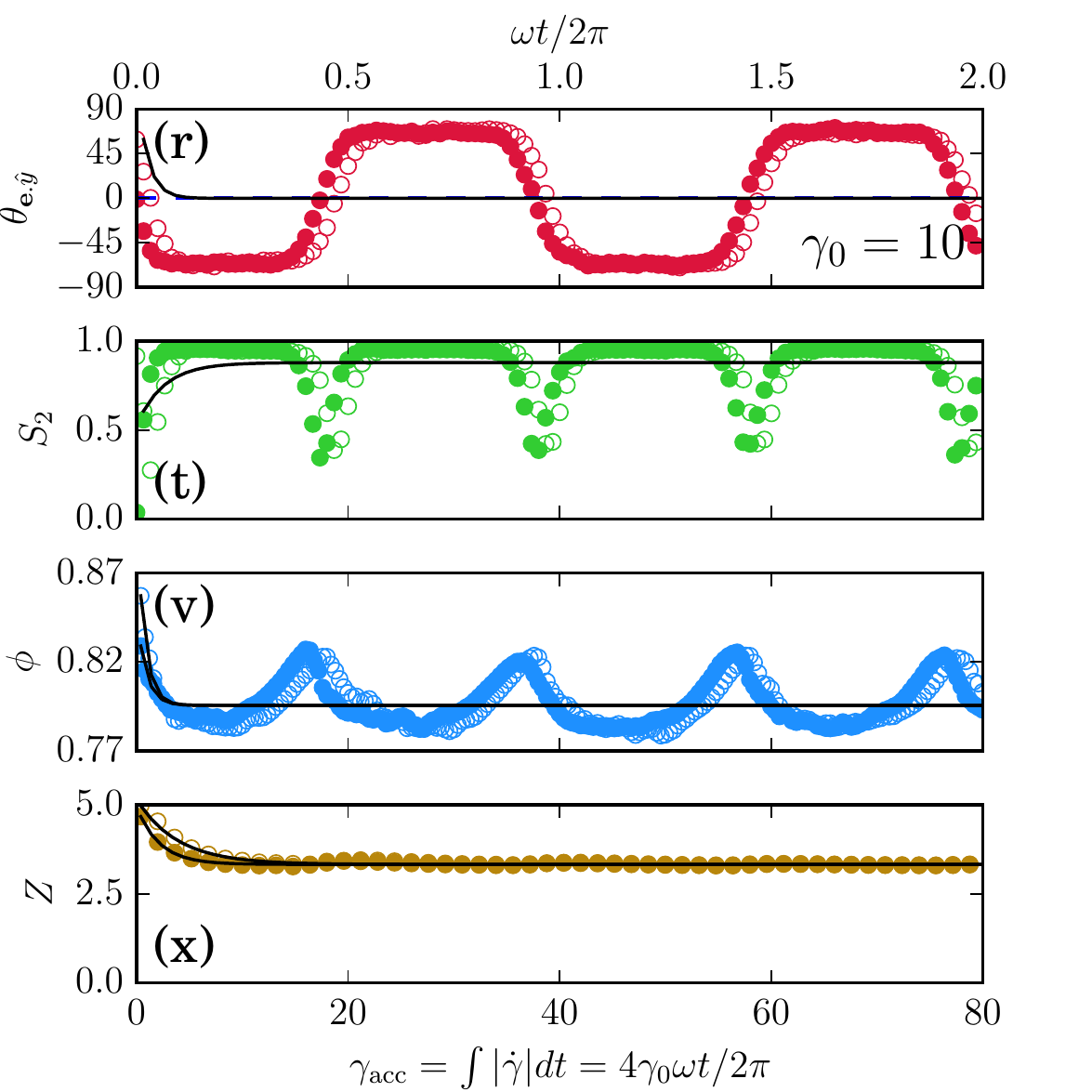}% Here is how to import EPS art

    \caption{\label{Evol} Evolution of the direction angle $\theta_{\bold e \cdot \hat{y}}$($^{\circ}$), the nematic order $S_2$, the packing fraction $\phi$, and the number of contacts $Z$ for the pre-sheared (empty symbols) and non-directional (full symbols) preparations at various $\gamma_0$ and at $J'\simeq0.1$. $(a)$ to $(h)$ represent the data corresponding to $\gamma_0=0.01$, while $(l)$ to $(p)$ and $(q)$ to $(x)$ illustrate the results for $\gamma_0=1$ and $\gamma_0=10$, respectively. Figures on the left column belong to frictional ($\mu_p=0.4$) configurations and figures on the right column belong to frictionless ($\mu_p=0$) suspensions. Black lines are the best fits of the data using the relaxation function $A(\gamma_\mathrm{acc})=A_\infty+(A_0-A_\infty)\exp(-\kappa \gamma_\mathrm{acc})$, where $A$ can be either of $\theta_{\bold e \cdot \hat{y}}$, $S_2$, $\phi$ or $Z$, and $\kappa^{-1}$ a relaxation strain. Dashed blue lines in the time series of $\theta_{\bold e \cdot \hat{y}}$ show the zero line.} 
   
\end{figure*}

As complementary information to Fig. 1 and Fig. 3 in the main text, Fig.~\ref{Evol} shows time evolutions of the direction angle $\theta_{\bold e \cdot \hat{y}}$, the nematic ordering $S_2$, the packing fraction $\phi$ and the number of contacts $Z$ of the two preparation protocols: pre-sheared (open symbols) and random (full symbols), at various $\gamma_0$ and at $J'\simeq0.1$ and for two different friction coefficients, $\mu_p=0.4$ (figures on the left side) and $\mu_p=0$ (figures on the right side). From Fig.~\ref{Evol} we can see the dependence of the relaxations on the magnitude of the oscillatory strain $\gamma_0$ and the friction coefficient $\mu_p$. For frictional cases, we can detect a convergence behaviour for all four properties at various $\gamma_0$, with almost instantaneous relaxations for $\gamma_0=1$ ($(l), (k)$, $(o)$ and $(m)$) and $\gamma_0=10$ ($(q), (s)$, $(u)$ and $(w)$), yet very slow relaxations at $\gamma_0=0.01$, especially for $\theta_{\bold e \cdot \hat{y}}$ and $S_2$ corresponding to more than $1000$ oscillations ($(a), (c)$, $(e)$ and $(g)$). In the frictionless case on the other hand, we can not detect any measurable convergence rate for the two protocols different properties at $\gamma_0=0.01$ within our simulation window ($(b), (d)$, $(f)$ and $(h)$). Increasing $\gamma_0$ to $1$ and $10$, frictionless packings also start to relax to the non-directional (see $(j), (l), (n), (p), (r)$, $(t)$, $(v)$ and $(x)$).

%\end{Appendixes}

\clearpage
%\FloatBarrier 
\bibliography{Oscillation}% Produces the bibliography via BibTeX.

\providecommand{\noopsort}[1]{}\providecommand{\singleletter}[1]{#1}%
\begin{thebibliography}{39}%
\makeatletter
\providecommand \@ifxundefined [1]{%
 \@ifx{#1\undefined}
}%
\providecommand \@ifnum [1]{%
 \ifnum #1\expandafter \@firstoftwo
 \else \expandafter \@secondoftwo
 \fi
}%
\providecommand \@ifx [1]{%
 \ifx #1\expandafter \@firstoftwo
 \else \expandafter \@secondoftwo
 \fi
}%
\providecommand \natexlab [1]{#1}%
\providecommand \enquote  [1]{``#1''}%
\providecommand \bibnamefont  [1]{#1}%
\providecommand \bibfnamefont [1]{#1}%
\providecommand \citenamefont [1]{#1}%
\providecommand \href@noop [0]{\@secondoftwo}%
\providecommand \href [0]{\begingroup \@sanitize@url \@href}%
\providecommand \@href[1]{\@@startlink{#1}\@@href}%
\providecommand \@@href[1]{\endgroup#1\@@endlink}%
\providecommand \@sanitize@url [0]{\catcode `\\12\catcode `\$12\catcode
  `\&12\catcode `\#12\catcode `\^12\catcode `\_12\catcode `\%12\relax}%
\providecommand \@@startlink[1]{}%
\providecommand \@@endlink[0]{}%
\providecommand \url  [0]{\begingroup\@sanitize@url \@url }%
\providecommand \@url [1]{\endgroup\@href {#1}{\urlprefix }}%
\providecommand \urlprefix  [0]{URL }%
\providecommand \Eprint [0]{\href }%
\providecommand \doibase [0]{https://doi.org/}%
\providecommand \selectlanguage [0]{\@gobble}%
\providecommand \bibinfo  [0]{\@secondoftwo}%
\providecommand \bibfield  [0]{\@secondoftwo}%
\providecommand \translation [1]{[#1]}%
\providecommand \BibitemOpen [0]{}%
\providecommand \bibitemStop [0]{}%
\providecommand \bibitemNoStop [0]{.\EOS\space}%
\providecommand \EOS [0]{\spacefactor3000\relax}%
\providecommand \BibitemShut  [1]{\csname bibitem#1\endcsname}%
\let\auto@bib@innerbib\@empty
%</preamble>
\bibitem [{\citenamefont {Liu}\ \emph {et~al.}(2020)\citenamefont {Liu},
  \citenamefont {Kijanka},\ and\ \citenamefont {Urban}}]{liu2020acoustic}%
  \BibitemOpen
  \bibfield  {author} {\bibinfo {author} {\bibfnamefont {H.-C.}\ \bibnamefont
  {Liu}}, \bibinfo {author} {\bibfnamefont {P.}~\bibnamefont {Kijanka}},\ and\
  \bibinfo {author} {\bibfnamefont {M.~W.}\ \bibnamefont {Urban}},\ }\bibfield
  {title} {\bibinfo {title} {Acoustic radiation force optical coherence
  elastography for evaluating mechanical properties of soft condensed matters
  and its biological applications},\ }\href@noop {} {\bibfield  {journal}
  {\bibinfo  {journal} {Journal of Biophotonics}\ }\textbf {\bibinfo {volume}
  {13}},\ \bibinfo {pages} {e201960134} (\bibinfo {year} {2020})}\BibitemShut
  {NoStop}%
\bibitem [{\citenamefont {Kleman}\ and\ \citenamefont
  {Laverntovich}(2007)}]{kleman2007soft}%
  \BibitemOpen
  \bibfield  {author} {\bibinfo {author} {\bibfnamefont {M.}~\bibnamefont
  {Kleman}}\ and\ \bibinfo {author} {\bibfnamefont {O.~D.}\ \bibnamefont
  {Laverntovich}},\ }\href@noop {} {\emph {\bibinfo {title} {Soft matter
  physics: an introduction}}}\ (\bibinfo  {publisher} {Springer Science \&
  Business Media},\ \bibinfo {year} {2007})\BibitemShut {NoStop}%
\bibitem [{\citenamefont {Ferry}(1980)}]{ferry1980viscoelastic}%
  \BibitemOpen
  \bibfield  {author} {\bibinfo {author} {\bibfnamefont {J.~D.}\ \bibnamefont
  {Ferry}},\ }\href@noop {} {\emph {\bibinfo {title} {Viscoelastic properties
  of polymers}}}\ (\bibinfo  {publisher} {John Wiley \& Sons},\ \bibinfo {year}
  {1980})\BibitemShut {NoStop}%
\bibitem [{\citenamefont {Dealy}\ and\ \citenamefont
  {Wissbrun}(2012)}]{dealy2012melt}%
  \BibitemOpen
  \bibfield  {author} {\bibinfo {author} {\bibfnamefont {J.~M.}\ \bibnamefont
  {Dealy}}\ and\ \bibinfo {author} {\bibfnamefont {K.~F.}\ \bibnamefont
  {Wissbrun}},\ }\href@noop {} {\emph {\bibinfo {title} {Melt rheology and its
  role in plastics processing: theory and applications}}}\ (\bibinfo
  {publisher} {Springer Science \& Business Media},\ \bibinfo {year}
  {2012})\BibitemShut {NoStop}%
\bibitem [{\citenamefont {Marenne}\ and\ \citenamefont
  {Morris}(2017)}]{marenne2017nonlinear}%
  \BibitemOpen
  \bibfield  {author} {\bibinfo {author} {\bibfnamefont {S.}~\bibnamefont
  {Marenne}}\ and\ \bibinfo {author} {\bibfnamefont {J.~F.}\ \bibnamefont
  {Morris}},\ }\bibfield  {title} {\bibinfo {title} {Nonlinear rheology of
  colloidal suspensions probed by oscillatory shear},\ }\href@noop {}
  {\bibfield  {journal} {\bibinfo  {journal} {Journal of Rheology}\ }\textbf
  {\bibinfo {volume} {61}},\ \bibinfo {pages} {797} (\bibinfo {year}
  {2017})}\BibitemShut {NoStop}%
\bibitem [{\citenamefont {Guazzelli}\ and\ \citenamefont
  {Pouliquen}(2018)}]{guazzelli2018rheology}%
  \BibitemOpen
  \bibfield  {author} {\bibinfo {author} {\bibfnamefont {{\'E}.}~\bibnamefont
  {Guazzelli}}\ and\ \bibinfo {author} {\bibfnamefont {O.}~\bibnamefont
  {Pouliquen}},\ }\bibfield  {title} {\bibinfo {title} {Rheology of dense
  granular suspensions},\ }\href@noop {} {\bibfield  {journal} {\bibinfo
  {journal} {Journal of Fluid Mechanics}\ }\textbf {\bibinfo {volume} {852}}
  (\bibinfo {year} {2018})}\BibitemShut {NoStop}%
\bibitem [{\citenamefont {Andreotti}\ \emph {et~al.}(2012)\citenamefont
  {Andreotti}, \citenamefont {Barrat},\ and\ \citenamefont
  {Heussinger}}]{andreotti2012shear}%
  \BibitemOpen
  \bibfield  {author} {\bibinfo {author} {\bibfnamefont {B.}~\bibnamefont
  {Andreotti}}, \bibinfo {author} {\bibfnamefont {J.-L.}\ \bibnamefont
  {Barrat}},\ and\ \bibinfo {author} {\bibfnamefont {C.}~\bibnamefont
  {Heussinger}},\ }\bibfield  {title} {\bibinfo {title} {Shear flow of
  non-brownian suspensions close to jamming},\ }\href@noop {} {\bibfield
  {journal} {\bibinfo  {journal} {Physical review letters}\ }\textbf {\bibinfo
  {volume} {109}},\ \bibinfo {pages} {105901} (\bibinfo {year}
  {2012})}\BibitemShut {NoStop}%
\bibitem [{\citenamefont {Salerno}\ \emph {et~al.}(2018)\citenamefont
  {Salerno}, \citenamefont {Bolintineanu}, \citenamefont {Grest}, \citenamefont
  {Lechman}, \citenamefont {Plimpton}, \citenamefont {Srivastava},\ and\
  \citenamefont {Silbert}}]{salerno2018effect}%
  \BibitemOpen
  \bibfield  {author} {\bibinfo {author} {\bibfnamefont {K.~M.}\ \bibnamefont
  {Salerno}}, \bibinfo {author} {\bibfnamefont {D.~S.}\ \bibnamefont
  {Bolintineanu}}, \bibinfo {author} {\bibfnamefont {G.~S.}\ \bibnamefont
  {Grest}}, \bibinfo {author} {\bibfnamefont {J.~B.}\ \bibnamefont {Lechman}},
  \bibinfo {author} {\bibfnamefont {S.~J.}\ \bibnamefont {Plimpton}}, \bibinfo
  {author} {\bibfnamefont {I.}~\bibnamefont {Srivastava}},\ and\ \bibinfo
  {author} {\bibfnamefont {L.~E.}\ \bibnamefont {Silbert}},\ }\bibfield
  {title} {\bibinfo {title} {Effect of shape and friction on the packing and
  flow of granular materials},\ }\href@noop {} {\bibfield  {journal} {\bibinfo
  {journal} {Physical Review E}\ }\textbf {\bibinfo {volume} {98}},\ \bibinfo
  {pages} {050901} (\bibinfo {year} {2018})}\BibitemShut {NoStop}%
\bibitem [{\citenamefont {Marschall}\ \emph {et~al.}(2019)\citenamefont
  {Marschall}, \citenamefont {Keta}, \citenamefont {Olsson},\ and\
  \citenamefont {Teitel}}]{marschall2019orientational}%
  \BibitemOpen
  \bibfield  {author} {\bibinfo {author} {\bibfnamefont {T.}~\bibnamefont
  {Marschall}}, \bibinfo {author} {\bibfnamefont {Y.-E.}\ \bibnamefont {Keta}},
  \bibinfo {author} {\bibfnamefont {P.}~\bibnamefont {Olsson}},\ and\ \bibinfo
  {author} {\bibfnamefont {S.}~\bibnamefont {Teitel}},\ }\bibfield  {title}
  {\bibinfo {title} {Orientational ordering in athermally sheared, aspherical,
  frictionless particles},\ }\href@noop {} {\bibfield  {journal} {\bibinfo
  {journal} {Physical review letters}\ }\textbf {\bibinfo {volume} {122}},\
  \bibinfo {pages} {188002} (\bibinfo {year} {2019})}\BibitemShut {NoStop}%
\bibitem [{\citenamefont {Nagy}\ \emph {et~al.}(2017)\citenamefont {Nagy},
  \citenamefont {Claudin}, \citenamefont {B{\"o}rzs{\"o}nyi},\ and\
  \citenamefont {Somfai}}]{nagy2017rheology}%
  \BibitemOpen
  \bibfield  {author} {\bibinfo {author} {\bibfnamefont {D.~B.}\ \bibnamefont
  {Nagy}}, \bibinfo {author} {\bibfnamefont {P.}~\bibnamefont {Claudin}},
  \bibinfo {author} {\bibfnamefont {T.}~\bibnamefont {B{\"o}rzs{\"o}nyi}},\
  and\ \bibinfo {author} {\bibfnamefont {E.}~\bibnamefont {Somfai}},\
  }\bibfield  {title} {\bibinfo {title} {Rheology of dense granular flows for
  elongated particles},\ }\href@noop {} {\bibfield  {journal} {\bibinfo
  {journal} {Physical Review E}\ }\textbf {\bibinfo {volume} {96}},\ \bibinfo
  {pages} {062903} (\bibinfo {year} {2017})}\BibitemShut {NoStop}%
\bibitem [{\citenamefont {Donev}\ \emph {et~al.}(2004)\citenamefont {Donev},
  \citenamefont {Cisse}, \citenamefont {Sachs}, \citenamefont {Variano},
  \citenamefont {Stillinger}, \citenamefont {Connelly}, \citenamefont
  {Torquato},\ and\ \citenamefont {Chaikin}}]{donev2004improving}%
  \BibitemOpen
  \bibfield  {author} {\bibinfo {author} {\bibfnamefont {A.}~\bibnamefont
  {Donev}}, \bibinfo {author} {\bibfnamefont {I.}~\bibnamefont {Cisse}},
  \bibinfo {author} {\bibfnamefont {D.}~\bibnamefont {Sachs}}, \bibinfo
  {author} {\bibfnamefont {E.~A.}\ \bibnamefont {Variano}}, \bibinfo {author}
  {\bibfnamefont {F.~H.}\ \bibnamefont {Stillinger}}, \bibinfo {author}
  {\bibfnamefont {R.}~\bibnamefont {Connelly}}, \bibinfo {author}
  {\bibfnamefont {S.}~\bibnamefont {Torquato}},\ and\ \bibinfo {author}
  {\bibfnamefont {P.~M.}\ \bibnamefont {Chaikin}},\ }\bibfield  {title}
  {\bibinfo {title} {Improving the density of jammed disordered packings using
  ellipsoids},\ }\href@noop {} {\bibfield  {journal} {\bibinfo  {journal}
  {Science}\ }\textbf {\bibinfo {volume} {303}},\ \bibinfo {pages} {990}
  (\bibinfo {year} {2004})}\BibitemShut {NoStop}%
\bibitem [{\citenamefont {Marschall}\ and\ \citenamefont
  {Teitel}(2019)}]{marschall2019shear}%
  \BibitemOpen
  \bibfield  {author} {\bibinfo {author} {\bibfnamefont {T.~A.}\ \bibnamefont
  {Marschall}}\ and\ \bibinfo {author} {\bibfnamefont {S.}~\bibnamefont
  {Teitel}},\ }\bibfield  {title} {\bibinfo {title} {Shear-driven flow of
  athermal, frictionless, spherocylinder suspensions in two dimensions: Stress,
  jamming, and contacts},\ }\href@noop {} {\bibfield  {journal} {\bibinfo
  {journal} {Physical Review E}\ }\textbf {\bibinfo {volume} {100}},\ \bibinfo
  {pages} {032906} (\bibinfo {year} {2019})}\BibitemShut {NoStop}%
\bibitem [{\citenamefont {Trulsson}(2018)}]{trulsson2018rheology}%
  \BibitemOpen
  \bibfield  {author} {\bibinfo {author} {\bibfnamefont {M.}~\bibnamefont
  {Trulsson}},\ }\bibfield  {title} {\bibinfo {title} {Rheology and shear
  jamming of frictional ellipses},\ }\href@noop {} {\bibfield  {journal}
  {\bibinfo  {journal} {Journal of Fluid Mechanics}\ }\textbf {\bibinfo
  {volume} {849}},\ \bibinfo {pages} {718} (\bibinfo {year}
  {2018})}\BibitemShut {NoStop}%
\bibitem [{\citenamefont {Az{\'e}ma}\ \emph {et~al.}(2015)\citenamefont
  {Az{\'e}ma}, \citenamefont {Radjai},\ and\ \citenamefont
  {Roux}}]{azema2015internal}%
  \BibitemOpen
  \bibfield  {author} {\bibinfo {author} {\bibfnamefont {{\'E}.}~\bibnamefont
  {Az{\'e}ma}}, \bibinfo {author} {\bibfnamefont {F.}~\bibnamefont {Radjai}},\
  and\ \bibinfo {author} {\bibfnamefont {J.-N.}\ \bibnamefont {Roux}},\
  }\bibfield  {title} {\bibinfo {title} {Internal friction and absence of
  dilatancy of packings of frictionless polygons},\ }\href@noop {} {\bibfield
  {journal} {\bibinfo  {journal} {Physical Review E}\ }\textbf {\bibinfo
  {volume} {91}},\ \bibinfo {pages} {010202} (\bibinfo {year}
  {2015})}\BibitemShut {NoStop}%
\bibitem [{\citenamefont {Brown}\ \emph {et~al.}(2011)\citenamefont {Brown},
  \citenamefont {Zhang}, \citenamefont {Forman}, \citenamefont {Maynor},
  \citenamefont {Betts}, \citenamefont {DeSimone},\ and\ \citenamefont
  {Jaeger}}]{brown2011shear}%
  \BibitemOpen
  \bibfield  {author} {\bibinfo {author} {\bibfnamefont {E.}~\bibnamefont
  {Brown}}, \bibinfo {author} {\bibfnamefont {H.}~\bibnamefont {Zhang}},
  \bibinfo {author} {\bibfnamefont {N.~A.}\ \bibnamefont {Forman}}, \bibinfo
  {author} {\bibfnamefont {B.~W.}\ \bibnamefont {Maynor}}, \bibinfo {author}
  {\bibfnamefont {D.~E.}\ \bibnamefont {Betts}}, \bibinfo {author}
  {\bibfnamefont {J.~M.}\ \bibnamefont {DeSimone}},\ and\ \bibinfo {author}
  {\bibfnamefont {H.~M.}\ \bibnamefont {Jaeger}},\ }\bibfield  {title}
  {\bibinfo {title} {Shear thickening and jamming in densely packed suspensions
  of different particle shapes},\ }\href@noop {} {\bibfield  {journal}
  {\bibinfo  {journal} {Physical Review E}\ }\textbf {\bibinfo {volume} {84}},\
  \bibinfo {pages} {031408} (\bibinfo {year} {2011})}\BibitemShut {NoStop}%
\bibitem [{\citenamefont {Trulsson}\ \emph {et~al.}(2017)\citenamefont
  {Trulsson}, \citenamefont {DeGiuli},\ and\ \citenamefont
  {Wyart}}]{trulsson2017effect}%
  \BibitemOpen
  \bibfield  {author} {\bibinfo {author} {\bibfnamefont {M.}~\bibnamefont
  {Trulsson}}, \bibinfo {author} {\bibfnamefont {E.}~\bibnamefont {DeGiuli}},\
  and\ \bibinfo {author} {\bibfnamefont {M.}~\bibnamefont {Wyart}},\ }\bibfield
   {title} {\bibinfo {title} {Effect of friction on dense suspension flows of
  hard particles},\ }\href@noop {} {\bibfield  {journal} {\bibinfo  {journal}
  {Physical Review E}\ }\textbf {\bibinfo {volume} {95}},\ \bibinfo {pages}
  {012605} (\bibinfo {year} {2017})}\BibitemShut {NoStop}%
\bibitem [{\citenamefont {Silbert}(2010)}]{silbert2010jamming}%
  \BibitemOpen
  \bibfield  {author} {\bibinfo {author} {\bibfnamefont {L.~E.}\ \bibnamefont
  {Silbert}},\ }\bibfield  {title} {\bibinfo {title} {Jamming of frictional
  spheres and random loose packing},\ }\href@noop {} {\bibfield  {journal}
  {\bibinfo  {journal} {Soft Matter}\ }\textbf {\bibinfo {volume} {6}},\
  \bibinfo {pages} {2918} (\bibinfo {year} {2010})}\BibitemShut {NoStop}%
\bibitem [{\citenamefont {Seto}\ \emph {et~al.}(2013)\citenamefont {Seto},
  \citenamefont {Mari}, \citenamefont {Morris},\ and\ \citenamefont
  {Denn}}]{seto2013discontinuous}%
  \BibitemOpen
  \bibfield  {author} {\bibinfo {author} {\bibfnamefont {R.}~\bibnamefont
  {Seto}}, \bibinfo {author} {\bibfnamefont {R.}~\bibnamefont {Mari}}, \bibinfo
  {author} {\bibfnamefont {J.~F.}\ \bibnamefont {Morris}},\ and\ \bibinfo
  {author} {\bibfnamefont {M.~M.}\ \bibnamefont {Denn}},\ }\bibfield  {title}
  {\bibinfo {title} {Discontinuous shear thickening of frictional hard-sphere
  suspensions},\ }\href@noop {} {\bibfield  {journal} {\bibinfo  {journal}
  {Physical review letters}\ }\textbf {\bibinfo {volume} {111}},\ \bibinfo
  {pages} {218301} (\bibinfo {year} {2013})}\BibitemShut {NoStop}%
\bibitem [{\citenamefont {Dong}\ and\ \citenamefont
  {Trulsson}(2020{\natexlab{a}})}]{dong2020unifying}%
  \BibitemOpen
  \bibfield  {author} {\bibinfo {author} {\bibfnamefont {J.}~\bibnamefont
  {Dong}}\ and\ \bibinfo {author} {\bibfnamefont {M.}~\bibnamefont
  {Trulsson}},\ }\bibfield  {title} {\bibinfo {title} {Unifying viscous and
  inertial regimes of discontinuous shear thickening suspensions},\ }\href@noop
  {} {\bibfield  {journal} {\bibinfo  {journal} {Journal of Rheology}\ }\textbf
  {\bibinfo {volume} {64}},\ \bibinfo {pages} {255} (\bibinfo {year}
  {2020}{\natexlab{a}})}\BibitemShut {NoStop}%
\bibitem [{\citenamefont {Irani}\ \emph {et~al.}(2014)\citenamefont {Irani},
  \citenamefont {Chaudhuri},\ and\ \citenamefont
  {Heussinger}}]{irani2014impact}%
  \BibitemOpen
  \bibfield  {author} {\bibinfo {author} {\bibfnamefont {E.}~\bibnamefont
  {Irani}}, \bibinfo {author} {\bibfnamefont {P.}~\bibnamefont {Chaudhuri}},\
  and\ \bibinfo {author} {\bibfnamefont {C.}~\bibnamefont {Heussinger}},\
  }\bibfield  {title} {\bibinfo {title} {Impact of attractive interactions on
  the rheology of dense athermal particles},\ }\href@noop {} {\bibfield
  {journal} {\bibinfo  {journal} {Physical review letters}\ }\textbf {\bibinfo
  {volume} {112}},\ \bibinfo {pages} {188303} (\bibinfo {year}
  {2014})}\BibitemShut {NoStop}%
\bibitem [{\citenamefont {Berger}\ \emph {et~al.}(2016)\citenamefont {Berger},
  \citenamefont {Az{\'e}ma}, \citenamefont {Douce},\ and\ \citenamefont
  {Radjai}}]{berger2016scaling}%
  \BibitemOpen
  \bibfield  {author} {\bibinfo {author} {\bibfnamefont {N.}~\bibnamefont
  {Berger}}, \bibinfo {author} {\bibfnamefont {E.}~\bibnamefont {Az{\'e}ma}},
  \bibinfo {author} {\bibfnamefont {J.-F.}\ \bibnamefont {Douce}},\ and\
  \bibinfo {author} {\bibfnamefont {F.}~\bibnamefont {Radjai}},\ }\bibfield
  {title} {\bibinfo {title} {Scaling behaviour of cohesive granular flows},\
  }\href@noop {} {\bibfield  {journal} {\bibinfo  {journal} {EPL (Europhysics
  Letters)}\ }\textbf {\bibinfo {volume} {112}},\ \bibinfo {pages} {64004}
  (\bibinfo {year} {2016})}\BibitemShut {NoStop}%
\bibitem [{\citenamefont {Singh}\ \emph {et~al.}(2019)\citenamefont {Singh},
  \citenamefont {Pednekar}, \citenamefont {Chun}, \citenamefont {Denn},\ and\
  \citenamefont {Morris}}]{singh2019yielding}%
  \BibitemOpen
  \bibfield  {author} {\bibinfo {author} {\bibfnamefont {A.}~\bibnamefont
  {Singh}}, \bibinfo {author} {\bibfnamefont {S.}~\bibnamefont {Pednekar}},
  \bibinfo {author} {\bibfnamefont {J.}~\bibnamefont {Chun}}, \bibinfo {author}
  {\bibfnamefont {M.~M.}\ \bibnamefont {Denn}},\ and\ \bibinfo {author}
  {\bibfnamefont {J.~F.}\ \bibnamefont {Morris}},\ }\bibfield  {title}
  {\bibinfo {title} {From yielding to shear jamming in a cohesive frictional
  suspension},\ }\href@noop {} {\bibfield  {journal} {\bibinfo  {journal}
  {Physical review letters}\ }\textbf {\bibinfo {volume} {122}},\ \bibinfo
  {pages} {098004} (\bibinfo {year} {2019})}\BibitemShut {NoStop}%
\bibitem [{\citenamefont {Blanc}\ \emph {et~al.}(2011)\citenamefont {Blanc},
  \citenamefont {Peters},\ and\ \citenamefont {Lemaire}}]{blanc2011local}%
  \BibitemOpen
  \bibfield  {author} {\bibinfo {author} {\bibfnamefont {F.}~\bibnamefont
  {Blanc}}, \bibinfo {author} {\bibfnamefont {F.}~\bibnamefont {Peters}},\ and\
  \bibinfo {author} {\bibfnamefont {E.}~\bibnamefont {Lemaire}},\ }\bibfield
  {title} {\bibinfo {title} {Local transient rheological behavior of
  concentrated suspensions},\ }\href@noop {} {\bibfield  {journal} {\bibinfo
  {journal} {Journal of Rheology}\ }\textbf {\bibinfo {volume} {55}},\ \bibinfo
  {pages} {835} (\bibinfo {year} {2011})}\BibitemShut {NoStop}%
\bibitem [{\citenamefont {Peters}\ \emph {et~al.}(2016)\citenamefont {Peters},
  \citenamefont {Ghigliotti}, \citenamefont {Gallier}, \citenamefont {Blanc},
  \citenamefont {Lemaire},\ and\ \citenamefont {Lobry}}]{peters2016rheology}%
  \BibitemOpen
  \bibfield  {author} {\bibinfo {author} {\bibfnamefont {F.}~\bibnamefont
  {Peters}}, \bibinfo {author} {\bibfnamefont {G.}~\bibnamefont {Ghigliotti}},
  \bibinfo {author} {\bibfnamefont {S.}~\bibnamefont {Gallier}}, \bibinfo
  {author} {\bibfnamefont {F.}~\bibnamefont {Blanc}}, \bibinfo {author}
  {\bibfnamefont {E.}~\bibnamefont {Lemaire}},\ and\ \bibinfo {author}
  {\bibfnamefont {L.}~\bibnamefont {Lobry}},\ }\bibfield  {title} {\bibinfo
  {title} {Rheology of non-brownian suspensions of rough frictional particles
  under shear reversal: A numerical study},\ }\href@noop {} {\bibfield
  {journal} {\bibinfo  {journal} {Journal of rheology}\ }\textbf {\bibinfo
  {volume} {60}},\ \bibinfo {pages} {715} (\bibinfo {year} {2016})}\BibitemShut
  {NoStop}%
\bibitem [{\citenamefont {Ness}\ and\ \citenamefont {Sun}(2016)}]{ness2016two}%
  \BibitemOpen
  \bibfield  {author} {\bibinfo {author} {\bibfnamefont {C.}~\bibnamefont
  {Ness}}\ and\ \bibinfo {author} {\bibfnamefont {J.}~\bibnamefont {Sun}},\
  }\bibfield  {title} {\bibinfo {title} {Two-scale evolution during shear
  reversal in dense suspensions},\ }\href@noop {} {\bibfield  {journal}
  {\bibinfo  {journal} {Physical Review E}\ }\textbf {\bibinfo {volume} {93}},\
  \bibinfo {pages} {012604} (\bibinfo {year} {2016})}\BibitemShut {NoStop}%
\bibitem [{\citenamefont {Lin}\ \emph {et~al.}(2016)\citenamefont {Lin},
  \citenamefont {Ness}, \citenamefont {Cates}, \citenamefont {Sun},\ and\
  \citenamefont {Cohen}}]{lin2016tunable}%
  \BibitemOpen
  \bibfield  {author} {\bibinfo {author} {\bibfnamefont {N.~Y.}\ \bibnamefont
  {Lin}}, \bibinfo {author} {\bibfnamefont {C.}~\bibnamefont {Ness}}, \bibinfo
  {author} {\bibfnamefont {M.~E.}\ \bibnamefont {Cates}}, \bibinfo {author}
  {\bibfnamefont {J.}~\bibnamefont {Sun}},\ and\ \bibinfo {author}
  {\bibfnamefont {I.}~\bibnamefont {Cohen}},\ }\bibfield  {title} {\bibinfo
  {title} {Tunable shear thickening in suspensions},\ }\href@noop {} {\bibfield
   {journal} {\bibinfo  {journal} {Proceedings of the National Academy of
  Sciences}\ }\textbf {\bibinfo {volume} {113}},\ \bibinfo {pages} {10774}
  (\bibinfo {year} {2016})}\BibitemShut {NoStop}%
\bibitem [{\citenamefont {Ness}\ \emph {et~al.}(2017)\citenamefont {Ness},
  \citenamefont {Xing},\ and\ \citenamefont {Eiser}}]{ness2017oscillatory}%
  \BibitemOpen
  \bibfield  {author} {\bibinfo {author} {\bibfnamefont {C.}~\bibnamefont
  {Ness}}, \bibinfo {author} {\bibfnamefont {Z.}~\bibnamefont {Xing}},\ and\
  \bibinfo {author} {\bibfnamefont {E.}~\bibnamefont {Eiser}},\ }\bibfield
  {title} {\bibinfo {title} {Oscillatory rheology of dense, athermal
  suspensions of nearly hard spheres below the jamming point},\ }\href@noop {}
  {\bibfield  {journal} {\bibinfo  {journal} {Soft Matter}\ }\textbf {\bibinfo
  {volume} {13}},\ \bibinfo {pages} {3664} (\bibinfo {year}
  {2017})}\BibitemShut {NoStop}%
\bibitem [{\citenamefont {Ness}\ \emph {et~al.}(2018)\citenamefont {Ness},
  \citenamefont {Mari},\ and\ \citenamefont {Cates}}]{ness2018shaken}%
  \BibitemOpen
  \bibfield  {author} {\bibinfo {author} {\bibfnamefont {C.}~\bibnamefont
  {Ness}}, \bibinfo {author} {\bibfnamefont {R.}~\bibnamefont {Mari}},\ and\
  \bibinfo {author} {\bibfnamefont {M.~E.}\ \bibnamefont {Cates}},\ }\bibfield
  {title} {\bibinfo {title} {Shaken and stirred: Random organization reduces
  viscosity and dissipation in granular suspensions},\ }\href@noop {}
  {\bibfield  {journal} {\bibinfo  {journal} {Science advances}\ }\textbf
  {\bibinfo {volume} {4}},\ \bibinfo {pages} {eaar3296} (\bibinfo {year}
  {2018})}\BibitemShut {NoStop}%
\bibitem [{\citenamefont {Dong}\ and\ \citenamefont
  {Trulsson}(2020{\natexlab{b}})}]{dong2020transition}%
  \BibitemOpen
  \bibfield  {author} {\bibinfo {author} {\bibfnamefont {J.}~\bibnamefont
  {Dong}}\ and\ \bibinfo {author} {\bibfnamefont {M.}~\bibnamefont
  {Trulsson}},\ }\bibfield  {title} {\bibinfo {title} {Transition from steady
  shear to oscillatory shear rheology of dense suspensions},\ }\href@noop {}
  {\bibfield  {journal} {\bibinfo  {journal} {Physical Review E}\ }\textbf
  {\bibinfo {volume} {102}},\ \bibinfo {pages} {052605} (\bibinfo {year}
  {2020}{\natexlab{b}})}\BibitemShut {NoStop}%
\bibitem [{\citenamefont {Pine}\ \emph {et~al.}(2005)\citenamefont {Pine},
  \citenamefont {Gollub}, \citenamefont {Brady},\ and\ \citenamefont
  {Leshansky}}]{pine2005chaos}%
  \BibitemOpen
  \bibfield  {author} {\bibinfo {author} {\bibfnamefont {D.~J.}\ \bibnamefont
  {Pine}}, \bibinfo {author} {\bibfnamefont {J.~P.}\ \bibnamefont {Gollub}},
  \bibinfo {author} {\bibfnamefont {J.~F.}\ \bibnamefont {Brady}},\ and\
  \bibinfo {author} {\bibfnamefont {A.~M.}\ \bibnamefont {Leshansky}},\
  }\bibfield  {title} {\bibinfo {title} {Chaos and threshold for
  irreversibility in sheared suspensions},\ }\href@noop {} {\bibfield
  {journal} {\bibinfo  {journal} {Nature}\ }\textbf {\bibinfo {volume} {438}},\
  \bibinfo {pages} {997} (\bibinfo {year} {2005})}\BibitemShut {NoStop}%
\bibitem [{\citenamefont {Corte}\ \emph {et~al.}(2008)\citenamefont {Corte},
  \citenamefont {Chaikin}, \citenamefont {Gollub},\ and\ \citenamefont
  {Pine}}]{corte2008random}%
  \BibitemOpen
  \bibfield  {author} {\bibinfo {author} {\bibfnamefont {L.}~\bibnamefont
  {Corte}}, \bibinfo {author} {\bibfnamefont {P.~M.}\ \bibnamefont {Chaikin}},
  \bibinfo {author} {\bibfnamefont {J.~P.}\ \bibnamefont {Gollub}},\ and\
  \bibinfo {author} {\bibfnamefont {D.~J.}\ \bibnamefont {Pine}},\ }\bibfield
  {title} {\bibinfo {title} {Random organization in periodically driven
  systems},\ }\href@noop {} {\bibfield  {journal} {\bibinfo  {journal} {Nature
  Physics}\ }\textbf {\bibinfo {volume} {4}},\ \bibinfo {pages} {420} (\bibinfo
  {year} {2008})}\BibitemShut {NoStop}%
\bibitem [{\citenamefont {Chwang}\ and\ \citenamefont
  {Wu}(1975)}]{chwang1975hydromechanics}%
  \BibitemOpen
  \bibfield  {author} {\bibinfo {author} {\bibfnamefont {A.~T.}\ \bibnamefont
  {Chwang}}\ and\ \bibinfo {author} {\bibfnamefont {T.}~\bibnamefont {Wu}},\
  }\bibfield  {title} {\bibinfo {title} {Hydromechanics of low-reynolds-number
  flow. part 2. singularity method for stokes flows},\ }\href@noop {}
  {\bibfield  {journal} {\bibinfo  {journal} {Journal of Fluid mechanics}\
  }\textbf {\bibinfo {volume} {67}},\ \bibinfo {pages} {787} (\bibinfo {year}
  {1975})}\BibitemShut {NoStop}%
\bibitem [{\citenamefont {Datta}\ and\ \citenamefont
  {Srivastava}(1999)}]{datta1999stokes}%
  \BibitemOpen
  \bibfield  {author} {\bibinfo {author} {\bibfnamefont {S.}~\bibnamefont
  {Datta}}\ and\ \bibinfo {author} {\bibfnamefont {D.~K.}\ \bibnamefont
  {Srivastava}},\ }\bibfield  {title} {\bibinfo {title} {Stokes drag on axially
  symmetric bodies: a new approach},\ }in\ \href@noop {} {\emph {\bibinfo
  {booktitle} {Proceedings of the Indian Academy of Sciences-Mathematical
  Sciences}}},\ Vol.\ \bibinfo {volume} {109}\ (\bibinfo {organization}
  {Springer},\ \bibinfo {year} {1999})\ pp.\ \bibinfo {pages}
  {441--452}\BibitemShut {NoStop}%
\bibitem [{\citenamefont {Ishima}\ and\ \citenamefont
  {Hayakawa}(2020)}]{ishima2020scaling}%
  \BibitemOpen
  \bibfield  {author} {\bibinfo {author} {\bibfnamefont {D.}~\bibnamefont
  {Ishima}}\ and\ \bibinfo {author} {\bibfnamefont {H.}~\bibnamefont
  {Hayakawa}},\ }\bibfield  {title} {\bibinfo {title} {Scaling laws for
  frictional granular materials confined by constant pressure under oscillatory
  shear},\ }\href@noop {} {\bibfield  {journal} {\bibinfo  {journal} {Physical
  Review E}\ }\textbf {\bibinfo {volume} {101}},\ \bibinfo {pages} {042902}
  (\bibinfo {year} {2020})}\BibitemShut {NoStop}%
\bibitem [{\citenamefont {Alonso-Marroqu\'{\i}n}\ and\ \citenamefont
  {Herrmann}(2004)}]{Alonso2004rachet}%
  \BibitemOpen
  \bibfield  {author} {\bibinfo {author} {\bibfnamefont {F.}~\bibnamefont
  {Alonso-Marroqu\'{\i}n}}\ and\ \bibinfo {author} {\bibfnamefont {H.~J.}\
  \bibnamefont {Herrmann}},\ }\bibfield  {title} {\bibinfo {title} {Ratcheting
  of granular materials},\ }\href
  {https://doi.org/10.1103/PhysRevLett.92.054301} {\bibfield  {journal}
  {\bibinfo  {journal} {Phys. Rev. Lett.}\ }\textbf {\bibinfo {volume} {92}},\
  \bibinfo {pages} {054301} (\bibinfo {year} {2004})}\BibitemShut {NoStop}%
\bibitem [{\citenamefont {Dong}\ and\ \citenamefont
  {Trulsson}(2020{\natexlab{c}})}]{dong2020oscillatory}%
  \BibitemOpen
  \bibfield  {author} {\bibinfo {author} {\bibfnamefont {J.}~\bibnamefont
  {Dong}}\ and\ \bibinfo {author} {\bibfnamefont {M.}~\bibnamefont
  {Trulsson}},\ }\bibfield  {title} {\bibinfo {title} {Oscillatory shear flows
  of dense suspensions at imposed pressure: Rheology and micro-structure},\
  }\href@noop {} {\bibfield  {journal} {\bibinfo  {journal} {arXiv preprint
  arXiv:2011.13215}\ } (\bibinfo {year} {2020}{\natexlab{c}})}\BibitemShut
  {NoStop}%
\bibitem [{\citenamefont {Trulsson}(2021)}]{trulsson2021reverse}%
  \BibitemOpen
  \bibfield  {author} {\bibinfo {author} {\bibfnamefont {M.}~\bibnamefont
  {Trulsson}},\ }\bibfield  {title} {\bibinfo {title} {Directional
  shear-jamming},\ }\href@noop {} {\bibfield  {journal} {\bibinfo  {journal}
  {arXiv}\ ,\ \bibinfo {pages} {2103.11115}} (\bibinfo {year}
  {2021})}\BibitemShut {NoStop}%
\bibitem [{\citenamefont {Yuan}\ \emph {et~al.}(2021)\citenamefont {Yuan},
  \citenamefont {Xing}, \citenamefont {Zheng}, \citenamefont {Li},
  \citenamefont {Yuan}, \citenamefont {Zhang}, \citenamefont {Zeng},
  \citenamefont {Xia}, \citenamefont {Tong}, \citenamefont {Kob}, \citenamefont
  {Zhang},\ and\ \citenamefont {Wang}}]{Yuan2021edwards}%
  \BibitemOpen
  \bibfield  {author} {\bibinfo {author} {\bibfnamefont {Y.}~\bibnamefont
  {Yuan}}, \bibinfo {author} {\bibfnamefont {Y.}~\bibnamefont {Xing}}, \bibinfo
  {author} {\bibfnamefont {J.}~\bibnamefont {Zheng}}, \bibinfo {author}
  {\bibfnamefont {Z.}~\bibnamefont {Li}}, \bibinfo {author} {\bibfnamefont
  {H.}~\bibnamefont {Yuan}}, \bibinfo {author} {\bibfnamefont {S.}~\bibnamefont
  {Zhang}}, \bibinfo {author} {\bibfnamefont {Z.}~\bibnamefont {Zeng}},
  \bibinfo {author} {\bibfnamefont {C.}~\bibnamefont {Xia}}, \bibinfo {author}
  {\bibfnamefont {H.}~\bibnamefont {Tong}}, \bibinfo {author} {\bibfnamefont
  {W.}~\bibnamefont {Kob}}, \bibinfo {author} {\bibfnamefont {J.}~\bibnamefont
  {Zhang}},\ and\ \bibinfo {author} {\bibfnamefont {Y.}~\bibnamefont {Wang}},\
  }\bibfield  {title} {\bibinfo {title} {Experimental test of the edwards
  volume ensemble for tapped granular packings},\ }\href
  {https://doi.org/10.1103/PhysRevLett.127.018002} {\bibfield  {journal}
  {\bibinfo  {journal} {Phys. Rev. Lett.}\ }\textbf {\bibinfo {volume} {127}},\
  \bibinfo {pages} {018002} (\bibinfo {year} {2021})}\BibitemShut {NoStop}%
\bibitem [{\citenamefont {Baule}\ \emph {et~al.}(2018)\citenamefont {Baule},
  \citenamefont {Morone}, \citenamefont {Herrmann},\ and\ \citenamefont
  {Makse}}]{Baule2018edwards}%
  \BibitemOpen
  \bibfield  {author} {\bibinfo {author} {\bibfnamefont {A.}~\bibnamefont
  {Baule}}, \bibinfo {author} {\bibfnamefont {F.}~\bibnamefont {Morone}},
  \bibinfo {author} {\bibfnamefont {H.~J.}\ \bibnamefont {Herrmann}},\ and\
  \bibinfo {author} {\bibfnamefont {H.~A.}\ \bibnamefont {Makse}},\ }\bibfield
  {title} {\bibinfo {title} {Edwards statistical mechanics for jammed granular
  matter},\ }\href {https://doi.org/10.1103/RevModPhys.90.015006} {\bibfield
  {journal} {\bibinfo  {journal} {Rev. Mod. Phys.}\ }\textbf {\bibinfo {volume}
  {90}},\ \bibinfo {pages} {015006} (\bibinfo {year} {2018})}\BibitemShut
  {NoStop}%
\end{thebibliography}%

\end{document}